\documentclass[%
 reprint,
superscriptaddress,
 amsmath,amssymb,
 aps,
]{revtex4-2}

\usepackage{graphicx}
\usepackage{dcolumn}
\usepackage{bm}

\usepackage[caption=false]{subfig}

\usepackage[hidelinks]{hyperref} 
\usepackage{siunitx} 
\begin{document}

\title{Second harmonic generation under doubly resonant lattice plasmon excitation}

\author{Sebastian Beer}
\thanks{These authors equally contributed to this work.}
\affiliation{Friedrich Schiller University Jena, Institute of Applied Physics,  07745 Jena, Germany}
\author{Jeetendra Gour}
\thanks{These authors equally contributed to this work.}
\affiliation{Friedrich Schiller University Jena, Institute of Applied Physics,  07745 Jena, Germany}
\author{Alessandro Alberucci}
\affiliation{Friedrich Schiller University Jena, Institute of Applied Physics,  07745 Jena, Germany}
\author{Christin David}
\affiliation{Friedrich Schiller University Jena, Institute of Condensed Matter Theory and Optics,  07743 Jena, Germany}
\author{Stefan Nolte}
\affiliation{Friedrich Schiller University Jena, Institute of Applied Physics,  07745 Jena, Germany}
\affiliation{Fraunhofer Institute for Applied Optics and Precision Engineering IOF, 07745 Jena, Germany}
\author{Uwe Detlef Zeitner}
\affiliation{Friedrich Schiller University Jena, Institute of Applied Physics,  07745 Jena, Germany}
\affiliation{Fraunhofer Institute for Applied Optics and Precision Engineering IOF, 07745 Jena, Germany}
\thanks{ equally contributed to this work.}


\begin{abstract}
Second harmonic generation is enhanced at the surface lattice resonance in plasmonic nanoparticle arrays. 
We carried out a parametric investigation on two-dimensional lattices composed of gold nanobars where the centrosymmetry is broken at oblique incidence. 
We study the influence of the periodicity, the incidence angle and the direction of the linear input polarization on the second harmonic generation.
Excitation of the surface lattice resonance either at the fundamental or second harmonic wavelength, achieved by varying the incidence angle, enhance the conversion efficiency.
As a special case, we demonstrate that both the wavelengths can be simultaneously in resonance for a specific period of the lattice.
In this double resonant case, maximum second harmonic power is achieved.
\end{abstract}

\keywords{Nonlinear optics, second harmonic, plasmonics, surface lattice resonance, double resonance}

\maketitle

\textit{
\textsuperscript{\textcopyright} \href{https://opg.optica.org/oe/fulltext.cfm?uri=oe-30-22-40884&id=511118}{Opt. Express 30, 40884-40896 [2022]},
Optica Publishing Group. Users may use, reuse, and build upon the article, or use the article for text or data mining, so long as such uses are for non-commercial purposes and appropriate attribution is maintained. All other rights are reserved.}

\section{Introduction}
Thanks to the fast development in manufacturing of subwavelength structures in a controlled and reliable way, metamaterials allow to  manipulate the light-matter interaction beyond the possibilities of conventinal materials.
Plasmonic metasurfaces \cite{Yu:2014} are bidimensional structures consisting of periodically arranged metallic nanoparticles.
Plasmons are the collective oscillation of conduction band electrons in metals, and they have been largely studied due to intriguing properties, such as dramatic subwavelength localization and extreme electromagnetic field enhancement \cite{Maier:2007}.
The optical properties depend on the electromagnetic mode supported by the elemental nanoparticles (localized surface plasmon resonance, LSPR) and on the long-range coupling determined by the periodic arrangement (surface lattice resonance, SLR)\cite{Kravets:2018}.
SLRs can emerge when the scattered field of a single nanoparticle is in phase with the emitted fields from the surrounding nanoparticles, exciting a collective plasmonic wave propagating along the lattice.
Accordingly, the conditions for SLR are closely related to the presence of a diffraction mode propagating along the lattice, also known as Rayleigh anomaly (RA)\cite{Fano:1941,Hessel:1965}.
The spectral position of the SLR depends on the nanoparticle periodicity and the surrounding refractive index, but can be tuned with the angle of incidence. 
At the SLR the oscillation damping is partially compensated \cite{Kravets:2008,Kravets:2018}, thus tuning the interplay between LSPR and SLR enables to design ultra narrow resonances \cite{Zou:2004,Auguie:2008,Binalam:2021}.
The associated strong field enhancement indeed boosts nonlinear effects.
In 1985 Couatz \textit{et al.} showed that, as the incidence angle is changed, SHG undergoes strong enhancement in metallic mono-dimensional gratings when the impinging beam is resonant with the \textit{nonlocal surface plasmon} \cite{Coutaz:1985}, what is today known as SLR.
However, in the same structure the SHG is also enhanced when the SLR condition is fullfield at the second harmonic frequency \cite{Inchaussandague:2017}.
Similarly, for two-dimensional (2D) nanoparticle arrays the SHG is enhanced for the SLR at the fundamental frequency \cite{Czaplicki:2016, Hooper:2018} and at the second harmonic \cite{Michaeli:2017}.
The primary role of collective response yields surprising and counter-intuitive phenomena, such as larger SHG for less dense metallic gratings \cite{Linden:2012,Czaplicki:2018} or the strong dependence on the structure  of the unit cell \cite{Czaplicki:2013}.
Even beyond the linear regime, plasmonic metasurfaces provide a versatile way to shape light \cite{Keren-Zur:2018}.
The conversion process can be further improved by utilizing multiple-resonant nanoparticles, where the LSPR enhancement occurs simultaneously at the pump and signal wavelengths \cite{Celebrano:2015, Soun:2021, Yang:2017, Linnenbank:2016, Wen:2018,Zeng:2018,ShamsMousavi:2019}.
Recently, the relevance of double-resonance in the presence of SLRs along orthogonal directions has been stressed out theoretically \cite{Huttunen:2019} and was experimentally demonstrated \cite{Stolt:2022}. \\ 
In this paper, we perform a parametric investigation to study the influence of the SLR on the SHG, exploring various double resonant conditions, where the SLR of the fundamental and second harmonic are simultaneously excited.
Therefore, multiple 2D rectangular arrays composed of centrosymmetric gold nanobars on fused silica are fabricated, whereby just the lattice period  in one direction is varied while the LSPR value is kept constant.
We measured the SHG from these metasurfaces at different angles of incidence and linear polarization directions. The required non-centrosymmetry for SHG is provided by the left-right symmetry breaking associated with a tilted incident wavefront.
Relatively high average output powers are achieved, thus allowing the experimental investigation of the nonlinear effects using general purpose and cheap equipment such as CMOS cameras and compact spectrometers.
Finally, we model our experimental results using the nonlinear inverse scattering method \cite{Roke:2004}, which indeed connects the SHG to the the linear response of the structure at the two involved wavelengths \cite{Obrien:2015}.

\section{Sample fabrication and linear characterization}
\begin{figure}
    \centering
    \includegraphics[width=0.5\textwidth]{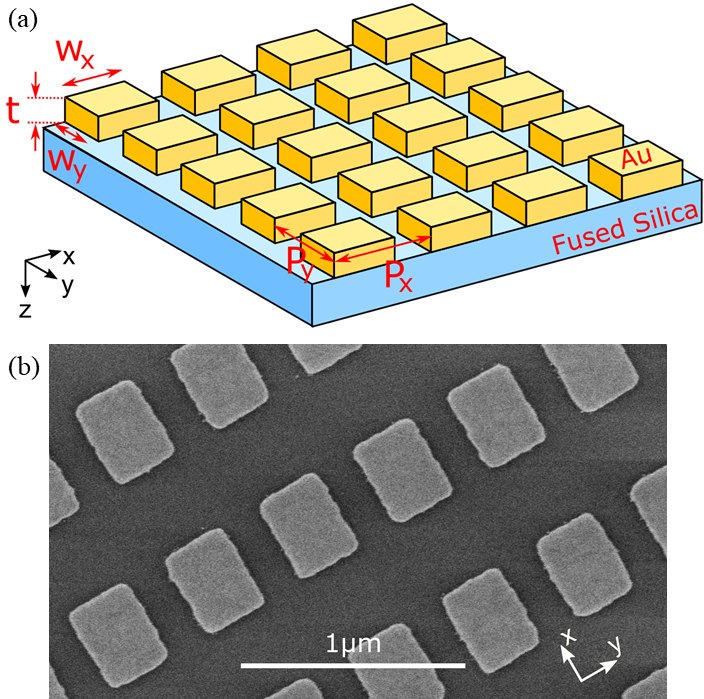}
    \caption[.]{
    (a) Sample design: 2D lattice of gold nanobars on a \SI{1}{\milli\meter} fused silica substrate. The geometrical parameters are $P_x\in \left[\SI{520}{\nano\meter},\  \SI{1200}{\nano\meter}\right]$, $P_y$=\SI{500}{\nano\meter}, $w_x$=\SI{400}{\nano\meter}, $w_y$=\SI{300}{\nano\meter}, $t$=\SI{50}{\nano\meter}. (b) SEM image of the sample featuring $P_x$=\SI{800}{\nano\meter}.
    }
    \label{fig:sample}
\end{figure}

We study several samples, where gold nanobars are arranged into 2D lattices on a \SI{1}{\milli\meter} thick fused silica substrate.
The metasurface template is shown in Fig.~\ref{fig:sample}a.
In all the fabricated samples, the shape of the bars working as nanoantennas is preserved. 
The elemental antenna is \SI{400}{\nano\meter} long ($w_x$), \SI{300}{\nano\meter} wide ($w_y$), and \SI{50}{\nano\meter} high.  
The period in the $y$-direction $P_y$ is fixed to \SI{500}{\nano\meter}.
We vary solely the period $P_x$ along the $x$-direction, the latter spanning the interval from 520 to \SI{1200}{\nano\meter} with steps of \SI{20}{\nano\meter}.
The structured area of each lattice is 3x\SI{3}{mm^2}. The wide area nanopatterning is accomplished using electron-beam lithography and standard metal lift-off technique 
(see Appendix~\ref{sec:fabrication}).
The scanning electron microscope (SEM) image in Fig.~\ref{fig:sample}b shows that our gold bars possess a large degree of uniformity, at the same time featuring very smooth edges on the nanometric scale.

To characterize the linear response of the fabricated metasurfaces, a supercontinuum laser is used as broadband light source in the setup (see Supplement~1, Fig.~S1), featuring a beneficial larger spatial coherence \cite{Droulias:2021, Binalam:2021}.
The light is weakly focused on the metasurface to a beam diameter of approx. \SI{300}{\micro\meter} with an achromatic $\SI{400}{\milli\meter}$ focal length lens. 
In all the experiments presented here the sample is rotated around the $y$-axis at the metasurface, changing the optical angle of incidence $\theta_{AOI}$ on the structure.
Thus, we are exploiting the angular dependence of the SLR only along the $x$-direction. Light propagates towards positive $z$-direction, and impinges the metasurface first before entering into the glass substrate.

\begin{figure}
    \centering
    \includegraphics[width=0.5\textwidth]{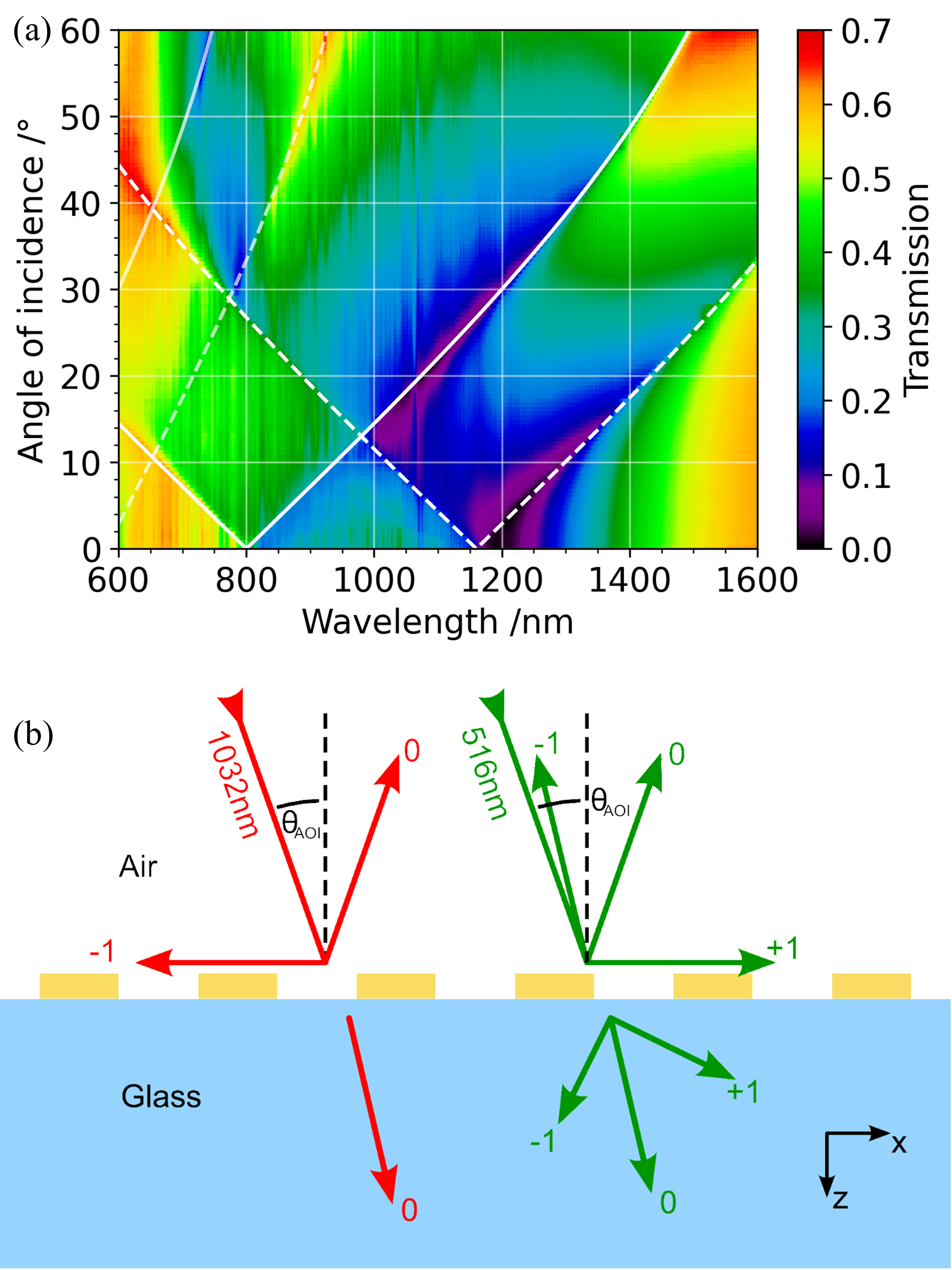}
    \caption{(a) Experimental linear transmission spectrum with TM-polarized light for the $P_x$=\SI{800}{\nano \meter} structure. Superposed white lines show the angle of the RAs at the (solid) air and (dashed) fused silica interfaces. The branches partially visible at shorter wavelengths stem from the RA at $|m|=2$. 
    (b) Sketch of the diffracted orders when passing through the structured interface with a period of $P_x$=\SI{770}{\nano\meter} and an incidence angle $\theta_{AOI}=\SI{19.5}{\degree}$ for  $\lambda=\SI{1032}{\nano\meter}$ and $\lambda = \SI{516}{\nano\meter}$.}
    \label{fig:800nmlinearspectra}
\end{figure}

The numerically computed LSPR for one isolated nanobar is at $\lambda$ =\SI{1276}{\nano\meter} for light polarized along the $x$-direction (TM or p-polarization) and along the $y$-direction (TE or s-polarization) at $\lambda$ =\SI{978}{\nano\meter} (see Supplement~1, Fig.~S2).
The corresponding bandwidth is very broad, approx. \SI{700}{\nano\meter}.
The experimental linear transmission spectrum as a function of the angle of incidence reveals the location of the plasmon resonances \cite{Kravets:2008}, exemplary shown for one lattice in Fig.~\ref{fig:800nmlinearspectra}a.
The broad transmission minimum at about $\lambda$ =\SI{1150}{\nano\meter} (for TE at $\lambda$ =\SI{900}{\nano\meter}, not shown here) emerges from the hybridization between LSPR and the $P_y$ lattice mode \cite{Khlopin:2017}.
Its position does not change with $P_x$ (see Supplement~1, Fig.~S3).
The narrow SLRs are visible as transmission anomalies \cite{Vecchi:2009,Humphrey:2014}, which closely follow the wavelength and the angle-depending RA conditions, represented by the superimposed white lines.
The incidence angle to achieve the RA condition either for a reflection grating (light still propagates in air) or transmission grating (light propagates in glass substrate) is given by \cite{Kravets:2008}:
\begin{equation} \label{eq:RA_air}
    \sin{\theta_{air}} = \pm 1 - \frac{m \lambda}{n_{air} P_x},
\end{equation}
\begin{equation} \label{eq:RA_glass}
  \sin{\theta_{glass}} = \pm \frac{n_{glass}}{n_{air}} - \frac{m \lambda}{n_{air} P_x},
\end{equation}
where $P_x$ is the period along $x$, $m$ the diffraction order, $\lambda$ the vacuum wavelength, and $n_{air}$ and $n_{glass}$ are the wavelength-dependent refractive indices of the air and fused silica medium, respectively.
At the RA, the diffraction order $m$ is diffracted by \SI{\pm 90}{\degree} and travels along the metasurface (see pictorial sketch in Fig.~\ref{fig:800nmlinearspectra}b).
The asymmetric index environment for the metasurface introduces two families of RA curves, thus introducing an additional degree of freedom to achieve the double resonance.
However, this comes at the price of a reduction of the SLR Q-factor \cite{Auguie:2010}. \\
\noindent
Simulated spectra stemming from RCWA (Rigorous Coupled Wave Analysis) 
match very well with the experimental results (see Supplement~1, Fig.~S4).
The good agreement shows that small technological imperfections (surface roughness, particle shape deformation, edge and corner rounding, etc.) have a negligible effect on the long distance interaction \cite{Auguie:2009} and that enough nanoparticles couple to each other to be approximated with an infinite array \cite{Rodriguez:2013}.\\
\begin{figure}
    \centering
    \includegraphics[width=.5\textwidth]{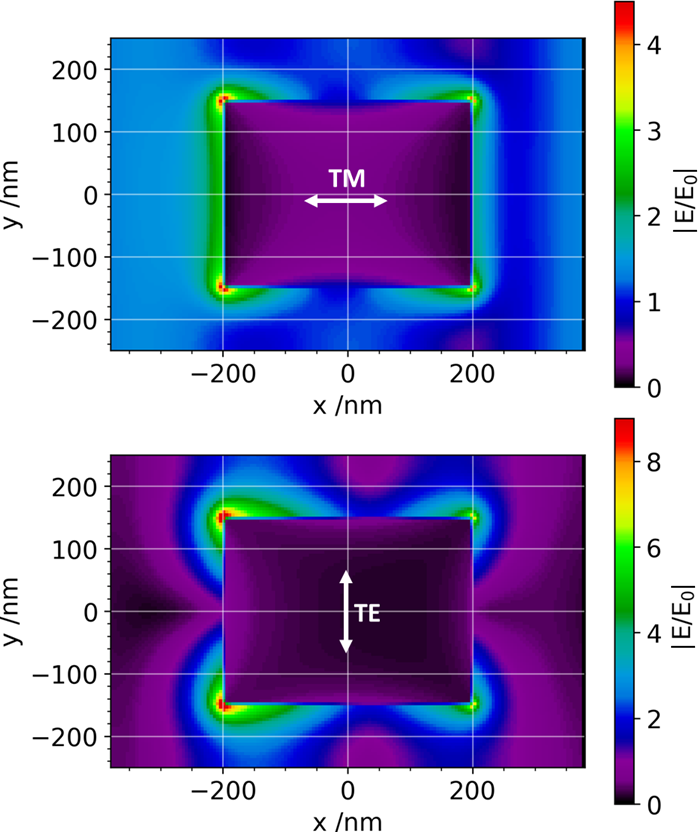}
    \caption{Simulated E-field amplitude in a unit cell plotted on the $xy$-plane at the nanobar mid-section when $P_x$=\SI{760}{\nano\meter}. The input beam at $\lambda$= \SI{1032}{\nano\meter}  satisfies the SLR condition being $\theta_{AOI}$= \SI{17}{\degree}. $E_0$ is the field amplitude in the absence of the metasurface.}
    \label{fig:E-field}
\end{figure}
Beyond the RA condition stemming from the momentum conservation, nonlinear effects are strongly dependent on the optical field distribution. 
We computed the optical field at the metasurface by using a commercial FDTD tool (see Appendix~\ref{sec:numerics}).
At normal incidence $\theta_{AOI}=\SI{0}{\degree}$, TM and TE polarized light excite an electric dipole in the nanobar along the $x$ and the $y$-direction, respectively.
A rotation of the metasurface around the $y$-axis leads to an asymmetric field distribution along $x$ (Fig.~\ref{fig:E-field}).
In the TE case, the dipole changes into an electric quadrupole distribution.
Note that the field enhancement is stronger for TE than for TM.
We confirmed the presence of a nonlocal plasmon oscillation with the field extending well beyond the unit cell \cite{Khlopin:2017} by measuring the power transported along $x$-direction on the metasurface plane depending on $\theta_{AOI}$.
Peaks and discontinuities in the power are observed whenever a SLR is excited. \\
We now focus on the two wavelengths corresponding to the fundamental (or pump, $\lambda_{FF}$=\SI{1032}{\nano\meter}) and its second harmonic ($\lambda_{SH}$=\SI{516}{\nano\meter}) in the frequency conversion experiment.
Figure~\ref{fig:characterization}a-d show the simulated linear transmission at $\lambda_{FF}$ (top row) and at $\lambda_{SH}$ (second row) versus $\theta_{AOI}$ and $P_x$, for TM (left column) and TE (right column) input polarizations.
Dips, peaks and jumps are mostly visible in proximity of the RA conditions, confirming that they originate from the SLR.
For TE polarized light, the SLR in air is prevented by the reflection of the substrate \cite{Auguie:2010}.
Due to the smaller fill factor, the absolute transmission increases for longer periods.
The average transmission at $\lambda_{FF}$ is lower than at $\lambda_{SH}$ due to the shorter spectral distance from the LSPR.\\

\newpage
\section{Experiments in the nonlinear regime}
\subsection{Second harmonic generation}

\begin{figure*}
\includegraphics[width=1.0\textwidth]{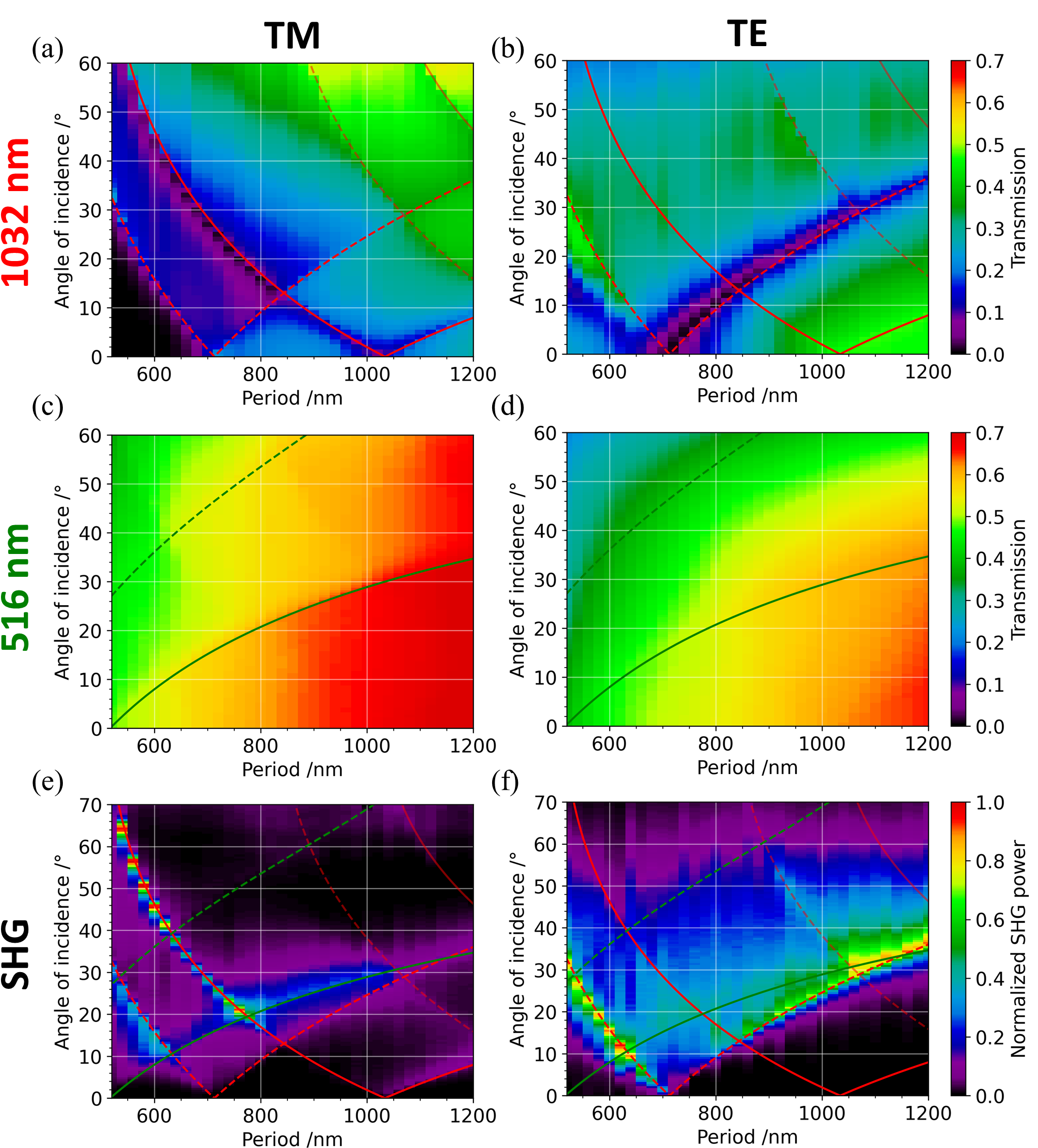}
  \caption{(a-d) Simulated linear transmission in the 0\textsuperscript{th} diffraction order for different incidence angles $\theta_{AOI}$ and periods $P_x$ for the (a-b) pump (\SI{1032}{\nano\meter}) and the (c-d) second harmonic wavelength (\SI{516}{\nano\meter}), for (left) TM and (right) TE-polarized pump light respectively. 
  (e-f) SHG power measured on the transmitted 0\textsuperscript{th} diffraction order normalized with respect to the absolute maximum in the plot. The superposed lines show conditions where the RA for $|m|=1$ is fulfilled (solid) in air and (dashed) in fused silica; the corresponding wavelengths are \SI{1032}{\nano\meter} (red)  and \SI{516}{\nano\meter} (green). 
  The branches partially visible at longer periods stem from the RA at $ |m| = 2$.}
  \label{fig:characterization}
\end{figure*}

For the nonlinear investigations we used an ultrafast laser delivering \SI{200}{\femto\second} (FWHM) short pulses at a wavelength of \SI{1032}{\nano\meter} with a repetition rate of \SI{200}{\kilo\hertz} and a variable average power.
The linearly polarized beam is focused onto the structured area to a \SI{250}{\micro\meter} beam diameter with a \SI{400}{\milli\meter} focal length lens.
The metasurface leads to a frequency conversion of the impinging light.
We investigate the SHG light which propagates in the direction of the incidence light (0\textsuperscript{th} transmission order).
We experimentally verified that SHG from the pure glass surface \cite{Rodriguez:2008} is not detectable in the range of powers we employ (more information see Appendix~\ref{sec:experiments} and Supplement~1).\\
While the centrosymmetry of gold is broken at the nanoparticle surface, the SHG is still prohibited due to the spatial symmetry. However, oblique incidence angles break the spatial symmetry and allow SHG from the structure \cite{Cirac:2012}, see Fig.~\ref{fig:E-field}.
The angular dependency of the SHG signal for different periods $P_x$ is shown in Fig.~\ref{fig:characterization}e-f.
By a direct comparison with the linear response (Fig.~\ref{fig:characterization}a-d), the strict connection between linear and nonlinear regime becomes immediately clear \cite{Michaeli:2017}.
At the RAs, a fast change versus the incident angle is observed in the SHG signal, including even the 2\textsuperscript{nd} diffraction order.
It is obvious that the SLRs have a significant impact on the frequency conversion.
Pronounced SLR related anomalies in the linear transmission also cause higher SHG efficiency.
The strong linear dependence on the input polarization is transferred to the SHG.
In fact, the TE case is mostly sensitive to SLR in glass (dashed red line), whereas for TM inputs SLRs are excited on both surfaces.
While lattice resonances of $\lambda_{FF}$  are more relevant in determining the SHG, 
surprisingly even the barely visible $\lambda_{SH}$ SLR in the linear regime has an influence on the frequency conversion strength for both input polarizations.\\ \noindent
\begin{figure*}
    \centering
    \includegraphics[width=0.95\textwidth]{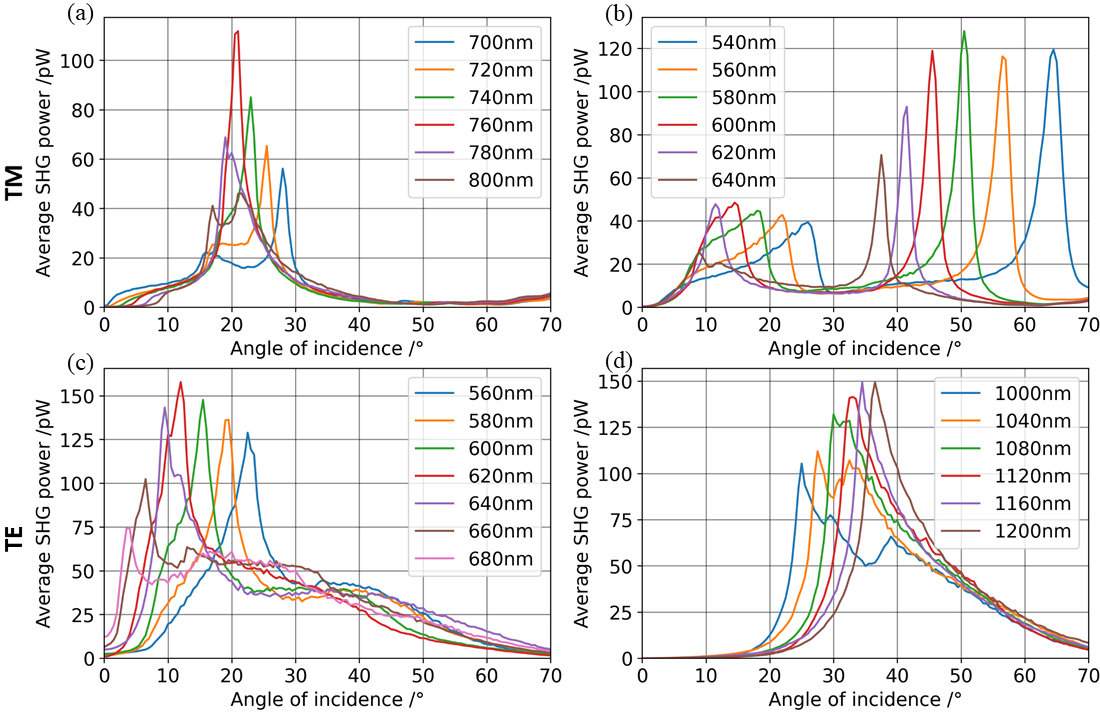}
    \caption{Measured average second harmonic power versus the incidence angle $\theta_{AOI}$ for different periods (different colored curves in each panel) and input polarizations (different panels).}
    \label{fig:SHG_cutout}
\end{figure*}
\noindent
Especially interesting are the multiple intersections between the $\lambda_{FF}$ and $\lambda_{SH}$ SLRs, corresponding to the double resonance case.
To see the behavior approaching the double resonance in more detail, we plotted in Fig.~\ref{fig:SHG_cutout} the average SHG power versus the incidence angle for the period region of interest.
For TM, the most prominent peak is at ($P_x$=\SI{760}{\nano\meter}, $\theta_{AOI}$=\SI{20.5}{\degree}) see Fig.~\ref{fig:characterization}e and Fig.~\ref{fig:SHG_cutout}a, corresponding to the double resonance.
Close to this condition, the linear diffracted light in air travels for both wavelengths along the nanoparticle array, schematically shown in Fig.~\ref{fig:800nmlinearspectra}b.
For slightly detuned periods, the two lattice resonances are clearly visible.
The region from $\SI{520}{nm}\leq P_x\leq \SI{620}{nm}$, $\theta_{AOI} < \SI{32}{\degree}$ (Fig.~\ref{fig:characterization}e) shows a SHG enhancement with a peak at the SLR for $\lambda_{FF}$ in glass.
For increasing periods the SHG peak undergoes a sudden increase reaching a local maximum in $P_x=\SI{620}{\nano\meter}, \theta_{AOI}=\SI{11.5}{\degree}$ (Fig.~\ref{fig:SHG_cutout}b), corresponding to the double resonance with $\lambda_{SH}$ in air and $\lambda_{FF}$ in glass.
The other double resonance of these two SLR at $(P_x=\SI{1140}{\nano\meter}, \theta_{AOI}=\SI{33.5}{\degree})$ leads also to a small signal peak (not shown).
On the other hand, the double resonance with the $\lambda_{SH}$ SLR in glass as well as the simultaneous excitation of both $\lambda_{FF}$ SLRs does not show an enhancement.
In the TE case for $\SI{520}{nm}<P_x<\SI{700}{nm}$ (Fig.~\ref{fig:characterization}f), the SHG rises for increasing periods, reaching a global maximum in $P_x=\SI{620}{\nano\meter}, \theta_{AOI}=\SI{11.5}{\degree}$ (Fig.~\ref{fig:SHG_cutout}c).
The effect of the double SLR in glass manifests as a fast decay of the peak for longer periods.
In the TE case there is another enhancement region $P_x\geq\SI{1080}{\nano\meter}$ (Fig.~\ref{fig:characterization}f), where two very close SLR peaks merge into a single peak for several periods after crossing with a third resonance (second order at $\lambda_{FF}$ in glass).
Other double resonances show minor effects:
near the fundamental SLR ($P_x=\SI{860}{\nano\meter}, \theta_{AOI}=\SI{13.5}{\degree}$), a relative SHG maximum is observed.
The TE polarized light provides higher conversion efficiencies than TM polarized light, related to the stronger field enhancement.
The highest conversion efficiency is achieved for TE polarized light at the double resonance.
On the other side, for TM inputs the strongest signal is observed at large angles and short periods along the SLR in air (Fig.~\ref{fig:characterization}e and \ref{fig:SHG_cutout}b), where the emitted SHG is slightly lower than in the double resonant case.
We ascribe the high efficiency to the larger E-field asymmetry and the longer coupling length between the incident beam and the metasurface.
We observed a saturation at large incidence angles $\theta_{AOI}>$\SI{45}{\degree}.
At normal incidence, the signal is usually below the detectable limit, except when the SLR is close to normal incidence, see Fig.~\ref{fig:characterization}f at $P_x\approx \SI{700}{\nano\meter}$.
This is probably caused by a weak quadrupolar mode \cite{Hooper:2018} or small fabrication imperfections \cite{Bachelier:2008} enhanced by the resonance.

An analyzer at the output is used to determine the polarization of the SHG.
Independently from the incident polarization, the SHG component along the $x$-direction (TM output) is always dominant.
Indeed, the SHG polarization is determined by the spatial symmetry breaking, which occurs along the $x$-direction due to the tilted input wavefront \cite{Cirac:2012}, see Fig.~\ref{fig:E-field}. 

\subsection{SHG characterization}

\begin{figure}
    \centering
    \includegraphics[width=.45\textwidth]{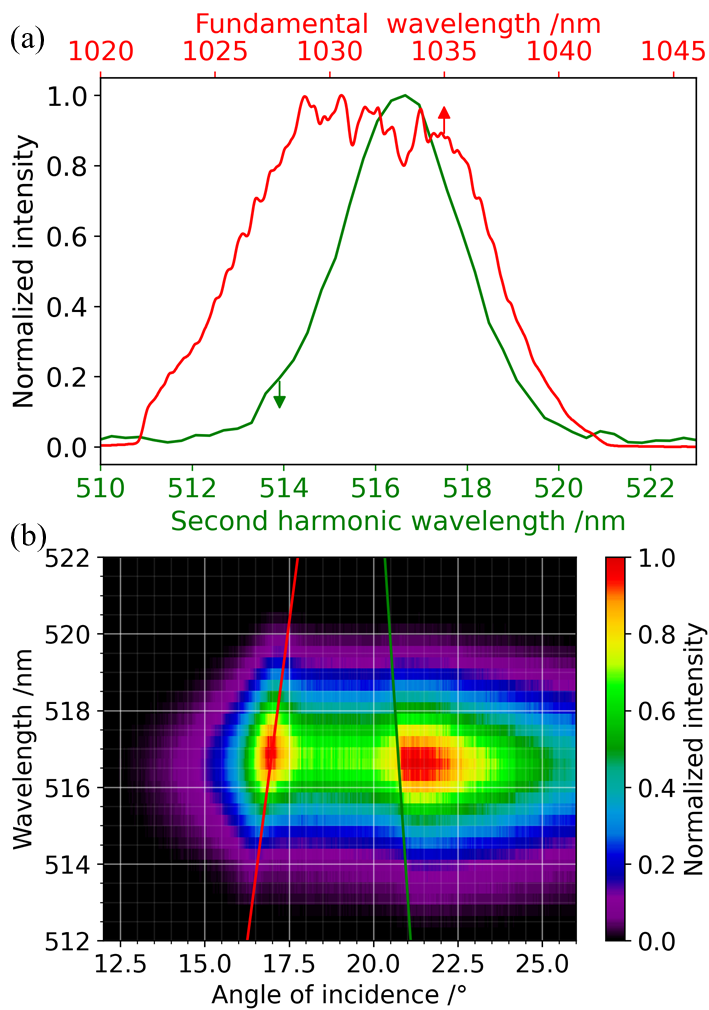}
    \caption{(a) Normalized spectra of the pump beam (red curve, top horizontal axis) and the corresponding SHG (green curve, bottom horizontal axis) at $P_x=\SI{760}{\nano\meter}$ and $\theta_{AOI}=\SI{20}{\degree}$. (b) Normalized SHG spectrum for $P_x$=\SI{800}{\nano\meter} at different angles of incidence $\theta_{AOI}$; the superposed solid lines show the RA for the fundamental (red color) and the second harmonic (green color).}
    \label{fig:SHG_spectrum}
\end{figure}

To verify that the origin of the detected signal is a second order nonlinear process, the emitted spectrum is analyzed via a compact back-thinned CCD spectrometer.
In the bandpass filter limited range from \SI{500}{\nano\meter} to \SI{789}{\nano\meter}, only a narrow peak is observed around the second harmonic, as shown in Fig.~\ref{fig:SHG_spectrum}a. Thus, two-photon luminescence is negligible in our case \cite{Boyd:1986}.
The center wavelength of the SHG (\SI{516.6}{\nano\meter}) is slightly red-shifted with respect to the half excitation wavelength, probably due to the asymmetry between forward and backward scattered light \cite{Han:2020}. Furthermore, the central wavelength of the SHG does not change noticeably with the incidence angle or the lattice period $P_x$ (not shown here).
A typical angular dependence of the SHG spectrum is shown in Fig.~\ref{fig:SHG_spectrum}b.
Sharp discontinuities are clearly visible at the indicated RAs, their positions changing with both the angle $\theta_{AOI}$ and the emitted wavelength $\lambda_{SH}$.
Noticeably, a spectrally-resolved measurement permits to reveal the sharpness of the transition.
The sharp features get indeed dimmed when using a detector uncapable of distinguishing the different wavelengths.

\begin{figure}
    \centering
    \includegraphics[width=.5\textwidth]{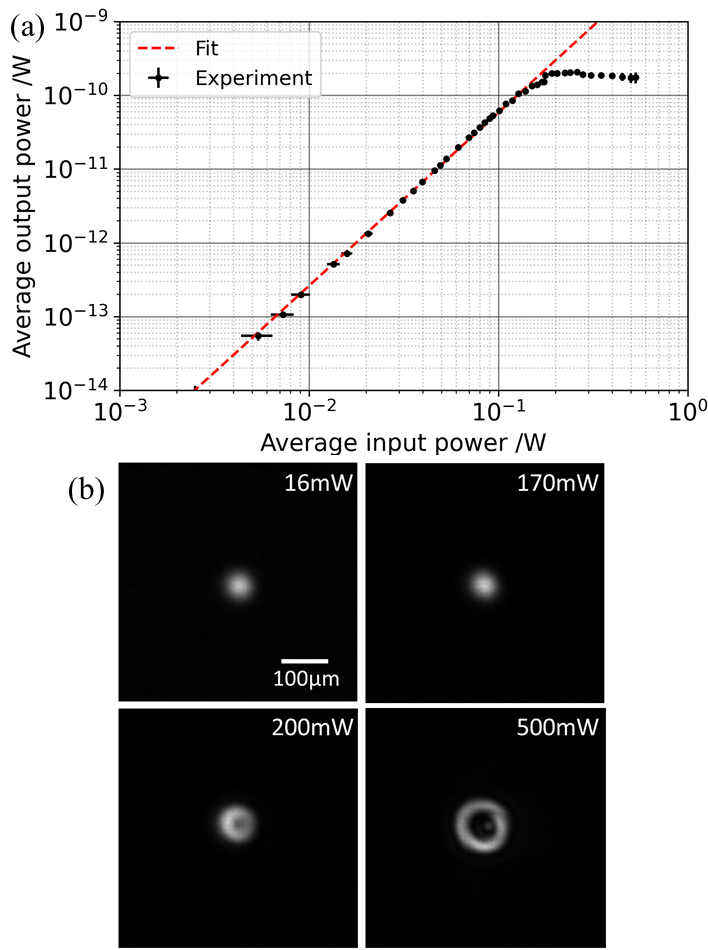}
  \caption{(a) Measured average SHG output power (black dots) and its fit (red dashed line) versus the average pump input power; a double logarithmic scale is used. Fitted power function: $P_{out} = c_0  P_{in}^{a}$, where $a=2.35$ and $c_0=1.30\cdot 10^{-8}\text{W}^{-1.35}$. (b) Pictures of the emitted beam at different pump powers. The gray scale is normalized to unity. Here $P_x=\SI{760}{\nano\meter}$, $\theta_{AOI}=\SI{21}{\degree}$, and the input is TM polarized.}
  \label{fig:SHG_power}
\end{figure}

The origin of the SHG signal is also verified by measuring the input-output power, as a double-log plot in Fig.~\ref{fig:SHG_power}a for the case of TM polarization and double resonance.
The output power ramps up with the average input power, reaching a threshold around $\SI{180}{\milli\watt}$ (corresponding to a peak intensity of \SI{\approx 15}{\giga\watt\per\square\centi\meter}) where the sample gets permanently damaged.
The maximum average output power is \SI{190}{\pico\watt}, corresponding to a maximum SHG conversion efficiency of \SI{1.06e-9}{}.
Taking the metasurface thickness into account, the maximum second order nonlinearity is around \SI{1.0}{\pico\meter\per\volt} (for more details, see Supplement~1), which is of the same order of magnitude with respect to similar plasmonic metasurfaces \cite{Stolt:2022, Inchaussandague:2017,Klein:2007}. 
Polynomial best-fitting finds a linear slope in the double log scale of \SI{2.35}{}. The exact value of the slope depends on the incidence angle and on the period.
Slightly steeper slopes than 2 have been already reported for nonlinear processes in metallic nanostructures \cite{Park:2012, Stolt:2022, Chen:2019}. 
Such a deviation from a pure square dependence can originate from the spatial and temporal nonlocality associated with the excitation of electrons in an energetic band \cite{Sun:1994}. 
We finally investigated what happens when the sample is damaged.
The recorded images in Fig.~\ref{fig:SHG_power}b reveal that above \SI{180}{\milli\watt} average input power a black area is formed in the center of the SHG spot, where the nano-structure is permanently modified.
Increasing the excitation power above this threshold leads to a wider modified area but also to a larger SHG spot. 
The output signal thus gradually decreases above the permanent damage threshold.

\newpage
\section{Interpretation of the results}

The simplest way to describe SHG from metallic metasurfaces is the nonlinear scattering technique first described in Ref.~\cite{Roke:2004}. 
Using the Lorentz reciprocity theorem, the electric field at the second harmonic $\mathbf{E}_\mathrm{SH}$ in the far field is given by the following surface integral extended to all the gold interfaces  \cite{Obrien:2015} 
\begin{equation} \label{eq:inverse_NL}
 \mathbf{E}_\mathrm{SH}(2\omega)\cdot \hat{l}\propto \gamma\int{\mathbf{P}^\mathrm{NL}(2\omega)\cdot \mathbf{E}_\mathrm{PW}(2\omega;l)\ dS},
\end{equation}
where ${P}^\mathrm{NL}_i(2\omega)=\sum_{jk}\chi^{(2)}_{ijk}E_j(\omega)E_k(\omega)\ (i,j,k=x,y,z)$  is the nonlinear polarization at $2\omega$, supposed to be non-vanishing only at the six interfaces enclosing each nanobar. 
The factor $\gamma$ is equal to $\cos\theta_{AOI}$ for $\hat{l}=\hat{x}$ (TM output) and equal to unity for $\hat{l}=\hat{y}$ (TE output).
Given we are interested in the transmitted SHG, the quantity $\mathbf{E}_\mathrm{PW}$ is the electric field generated on the sample by a plane wave which propagates from the observation point (coincident with the direction in the far field corresponding to the $0^{th}$ order) to the metasurface  at $2\omega$ with polarization parallel to $\hat{l}$.
The nonlinear polarization $\mathbf{P}^\mathrm{NL}$ depends on the linear coupling of the fundamental beam  propagating in the $z$-direction and impinging on the sample, whereas $\mathbf{E}_\mathrm{PW}$ depends on the linear coupling at $2\omega$ for waves propagating towards negative $z$. According to Equation~\eqref{eq:inverse_NL}, the overlap integral between these two quantities determines the amount of generated SH (for more information, see the Supplement~1). 
SHG is then maximized in correspondence to the linear resonances at $\omega$ and 2$\omega$, in agreement with the coupled dipole approximation \cite{Michaeli:2017}, with resonance at the fundamental intervening two times in the case of Type I SHG. 
Equation~\eqref{eq:inverse_NL} permits to understand why in the experiments the maxima of SHG are observed always at some RA, but not all the RAs correspond to a peak of nonlinear emission (e.g. SLR in air for TE inputs) \cite{Celebrano:2015}. 
The SHG indeed depends also on the spatial overlap between the two fields in the linear regime, the overlap being moreover modulated by the nonlinear coefficient $\chi^{(2)}$ of the nonlinear lattice.
When comparing different lattices, maximum SHG is obtained when double resonance occurs, i.e., for periods where fundamental and second harmonic resonances are observed for the same incidence angle $\theta_{AOI}$.

A more physical interpretation can be given in correspondence to the lattice resonances. 
Owing to the momentum conservation, the pump beam efficiently excites a propagating delocalized surface plasmon at $\omega$. 
The second-order nonlinearity of the metasurface then generates a plasmon at 2$\omega$. 
The last step is the coupling of the plasmon at $2\omega$ with the free space: this is maximized when a RA in the linear regime is observed at 2$\omega$. \\ 
\begin{figure*}
    \includegraphics[width=\textwidth]{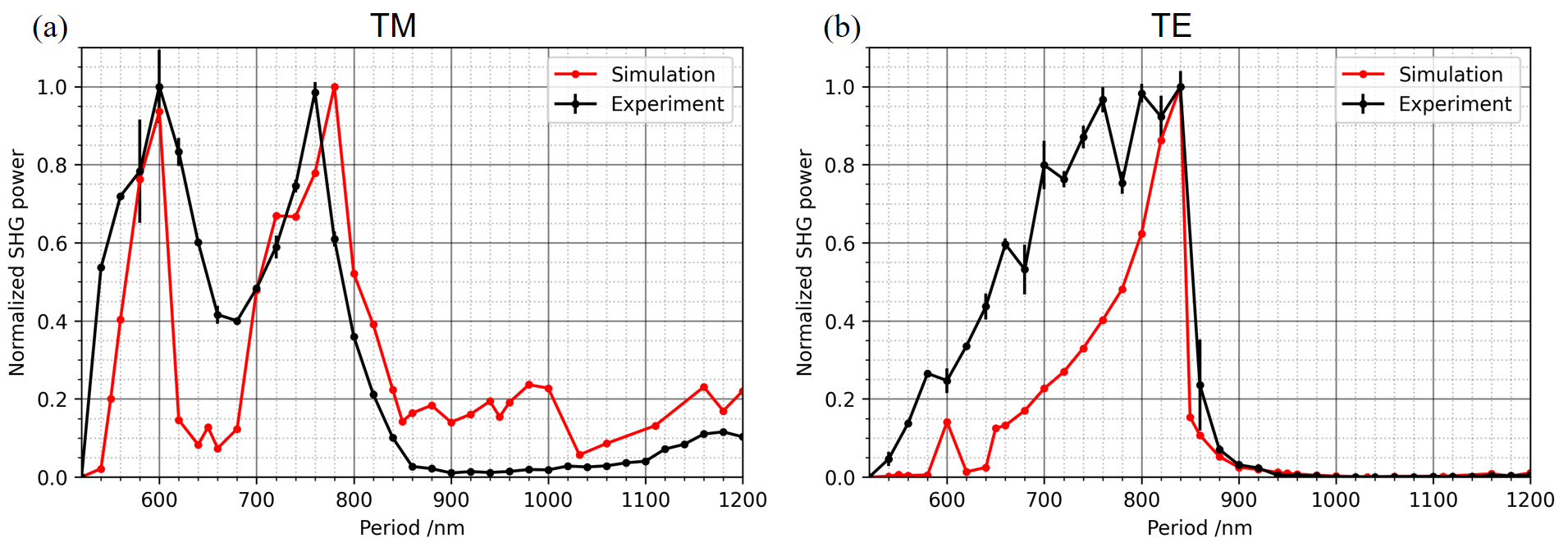}
  \caption{Experimental (black lines) and numerical (red lines) SHG power versus the period $P_x$ along a SLR at the fundamental frequency. The chosen incidence angle $\theta_{AOI}$ for each $P_x$ obeys the SLR condition for $\lambda_{FF}=\SI{1032}{\nano\meter}$ in air for $|m|=1$. Both the quantities are normalized to unity. Left and right panel correspond to TM and TE input, respectively.}
  \label{fig:nonlinear inverse scattering}
\end{figure*}
We finally compare the predictions of the nonlinear inverse scattering model with the measured data. 
Specifically, in Fig.~\ref{fig:nonlinear inverse scattering} the theoretical (red lines) and experimental (black lines) normalized SHG in correspondence to the calculated RA in air (Eq.~\ref{eq:RA_air}) for $|m|= 1$ and $\lambda_{FF}=\SI{1032}{\nano\meter}$.
For TM polarized pump (Fig.~\ref{fig:nonlinear inverse scattering}a), the experiment and simulation have a peak near the double resonant condition $P_x=\SI{760}{\nano\meter}$.
For lower periods the SHG power decreases, and the experiment features a lower depression than predicted.
The signal reaches another maximum at $P_x=\SI{600}{\nano\meter}$.
For even shorter periods and higher incident angles the signal suddenly drops because the calculated RA deviates from the SHG peak.
The experimental results in Fig.~\ref{fig:SHG_cutout}(b) show that the maximum signal stays constant for higher incidence angles. 
In the TE case plotted in panel Fig.~\ref{fig:nonlinear inverse scattering}(b), the experiment shows a broad peak due to the interaction between the double resonances at $P_x=\SI{760}{\nano\meter}$ and the crossing with the SLR of the pump wavelength in glass at $P_x=\SI{840}{\nano\meter}$. 
Interestingly, the latter does not cause any visible effect in the TM case.
The simulations just predict a narrower peak at $P_x=\SI{840}{\nano\meter}$, not the double resonance.
Limitation of the simulation is the prediction of the SHG enhancement at the second harmonic SLR.
Finally, we stress out that the numerical simulations also confirm that i) the main contribution of SHG is TM polarized, no matter what the polarization state of the pump is; ii) The SHG is stronger for TE than for TM polarization input.

\section{Conclusion}
We experimentally demonstrated the role of double surface lattice resonance in maximizing the SHG in plasmonic metasufaces. 
Our parametric study on centro-symmetric 2D nanobar arrays confirms \cite{Michaeli:2017} that local maxima of the SHG occur in correspondence to the SLR either at $\omega$ or 2$\omega$, thus
leading to a further enhancement of frequency conversion when both pump and emitted SLR get excited simultaneously.
Such a result has been predicted based upon the discrete dipole approximations \cite{Huttunen:2019} and recently experimentally confirmed by tilting the sample along two orthogonal directions \cite{Stolt:2022}.
Here we created the double resonance by tilting the metasurface with a specific period in just one direction.
The close connection between the linear and the nonlinear response of the metasurface has been theoretically confirmed using the nonlinear inverse scattering approach, providing a generalization of the Miller's rule to the SLR case \cite{Obrien:2015}.
Finally, due to the high quality of our samples and the fact that we are exploiting the double resonance, we achieved a maximum SHG conversion efficiency of \SI{1.06e-9}{}.
Such a comparatively large efficiency, in combination with high average excitation power, allowed us the direct measurement of both the spatial profile and spectrum of the generated second harmonic.

Straightforward extensions of our work include nano-antennas of different shape, with specific attention to nano-gapped resonators. A deeper theoretical treatment will be pursued in order to fully understand the nonlinear coupling between the nonlocal surface plasmons propagating across the metasurface, for example concentrating on the role played by the multipolar response of the elemental antenna. \\

\section{Appendix}

\subsection{Fabrication}
\label{sec:fabrication}

To fabricate 2D gold nanobar arrays, we have used a standard metal lift-off process.
First, we took the clean fused silica substrate and pattern the nanostructures using electron-beam lithography on a  bi-layer of positive tone resists (\SI{300}{\nano\meter}  ARP617.06 and \SI{100}{\nano\meter} ARP6200.04) using a \SI{10}{\nano\meter} gold layer as a conductive surface on top.
The sample was then developed in \mbox{AR600-546} for \SI{30}{\second} to remove the ARP6200.04 resist layer.
The remaing {ARP617.06} resist was then developed in MIBK:IPA~(1:1) for \SI{80}{\second}.
A \SI{50}{\nano\meter} Au layer was then deposited using an electron beam evaporator at \SI{1}{\nano\meter\per\second} deposition rate at a pressure lower than \SI{2e-5}{\milli\bar}, using a \SI{3}{\nano\meter} adhesion layer made of titanium. 
After the deposition process, a lift-off based on acetone and isopropanol is performed to realize the gold nanobar arrays on the fused silica substrate.

\subsection{Numerical simulations}
\label{sec:numerics}
Linear optical transmission spectra of a 2D array of nanobars were simulated using a commercial Finite Difference Time Domain (FDTD) solver by Ansys Lumerical and Rigorous Coupled Wave Analysis (RCWA) \cite{Moharam:1983}.  
The gold  optical  constants  used  for  simulations were taken from Ref.~\cite{Yakubovsky:2017}.
Transmission computed from FDTD and RCWA results are in very good agreement with each other, regardless of the geometric parameters of the structure.
Due to the larger computation time required by the FDTD simulations, RCWA is a better choice for investigating the parametric dependence of the optical response of the nanostructures. 
In RCWA-based simulations, 225 spatial harmonics were enough to obtain a good match with the experimental results for the 2D grating. 
After the simulation, the reflection from the backside of the substrate was taken into account in order to be comparable with the experiment.
The map of the electric fields on the surface of the nanostructures is simulated using FDTD solver. FDTD solutions are also used in the computation of the overlap integral providing the SHG according to the nonlinear inverse scattering technique. 
In all FDTD simulations, Bloch boundary conditions were used in $x$ and $y$-directions, whereas PML (Perfectly Matched Layer) boundary conditions were applied along $z$. The nanobar object was meshed \SI{5}{\nano\meter} on the transverse plane $xy$-directions and \SI{2}{\nano\meter} on the $z$-direction using mesh override region.

\subsection{Nonlinear detection}\label{sec:experiments}
The optical intensity at the sample plane is imaged onto a standard CMOS camera to investigate the emitted SHG power.
Bandpass filters limit the detectable spectral range from \SI{500}{\nano\meter} to \SI{540}{\nano\meter}.
The signal detected on the camera is calibrated versus the SHG of a nonlinear crystal, losses in the system were taken into account.
The linear behavior of the camera in the range of interest is verified.
The validity of the camera measurement at low signal is also cross-checked with a photo-multiplier tube. 
For comparison we verified that the input-power dependency of the SHG from a bulk nonlinear crystal (BBO) show a quadratic behavior (slope 2.04). 
We use a camera because several SHG output beams, parallel to the original 0\textsuperscript{th} transmission order, can arise at high incidence angles and large periods, due to multiple reflections inside the glass substrate of the sample.
The camera allows an easy distinction of these few millimeters separated replicas.
In addition, the beam is laterally displaced when the whole substrate is rotated around the $y$-axis.
Finally, for better visibility we only showed the response for positive angles, but we scanned from \SI{-70}{\degree} to \SI{+70}{\degree} to ensure that the SHG response under the employed illumination conditions is equal for negative and positive angles within our experimental accuracy.
Further details regarding the experiment can be found in the Supplement~1.

\begin{acknowledgments}
The authors thank the Deutsche Forschungsgemeinschaft (DFG) for funding the project (project number 398816777) within the framework of the CRC 1375 NOA.
We also acknowledge the valuable support of Werner Rockstroh, Natali Sergeev, Detlef Schelle, Holger Schmidt and Daniel Voigt in the fabrication of the nano-structured samples.
\end{acknowledgments}

\newpage
\bibliography{references}

\clearpage
\pagebreak
\widetext
\begin{center}
\textbf{\large \underline{Supporting Information:} \vspace{0.3cm} \linebreak  Second harmonic generation under doubly resonant lattice plasmon excitation}
\end{center}
\setcounter{equation}{0}
\setcounter{figure}{0}
\setcounter{table}{0}
\setcounter{page}{1}
\setcounter{section}{0}
\makeatletter
\renewcommand{\theequation}{S\arabic{equation}}
\renewcommand{\thefigure}{S\arabic{figure}}
\renewcommand{\bibnumfmt}[1]{[S#1]}
\noindent
In this supplemental material we first provide more details on the linear response of the structure, including the setup and the comparison between experiments and numerical simulations. The second part contains more details about the setup we used for the measurement of the SHG, including how we calibrated our camera. In the last part we provide all the step in deriving the generalization of the nonlinear inverse scattering to the case of surface lattice resonances. Finally, the method to find the effective second-order nonlinear tensor is presented.

\section{Linear transmission spectrum}
\subsection{Linear transmission spectrum setup}
\begin{figure}[h]
\centering
\includegraphics[width=0.5\textwidth]{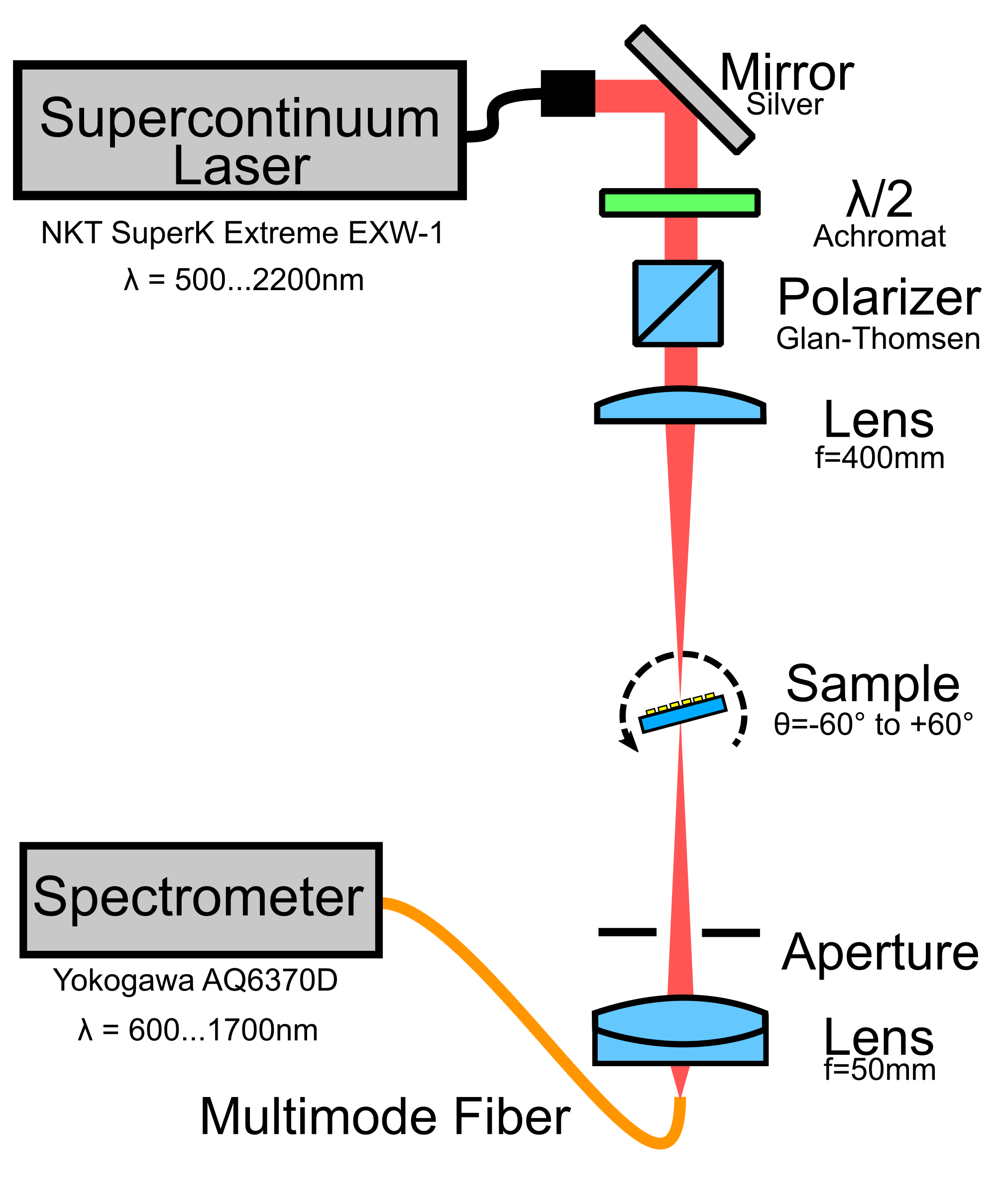}
  \caption{Linear transmission setup capturing the 0\textsuperscript{th} order.}
  \label{fig:linearsetup}
\end{figure}

\noindent
Figure~\ref{fig:linearsetup} shows the setup for the experimental linear transmission measurements. 
The beam from a supercontinuum laser (SuperK Extreme EXW-1 by NKT) is polarized, expanded to approx. \SI{4}{\milli\meter} and slightly focused onto the sample with a \SI{400}{\milli\meter} lens, similarly to the SHG characterization setup in Fig.~\ref{fig:SHGsetup}.
The transmitted 0\textsuperscript{th} order is coupled into a multimode fiber guiding it to a scanning spectrometer (AQ6370D by Yokogawa).
Changing the incidence angle leads to a plan-parallel beam displacement by the \SI{1}{\milli\meter} thick fused silica substrate, which also changes the coupling into the fiber.
This is taken into account by the reference measurements were performed on the unstructured area of the sample.

\newpage

\newpage

\subsection{Simulated extinction cross-section of an isolated nanobar}
\begin{figure}[h]
\centering
\includegraphics[width=0.6\textwidth]{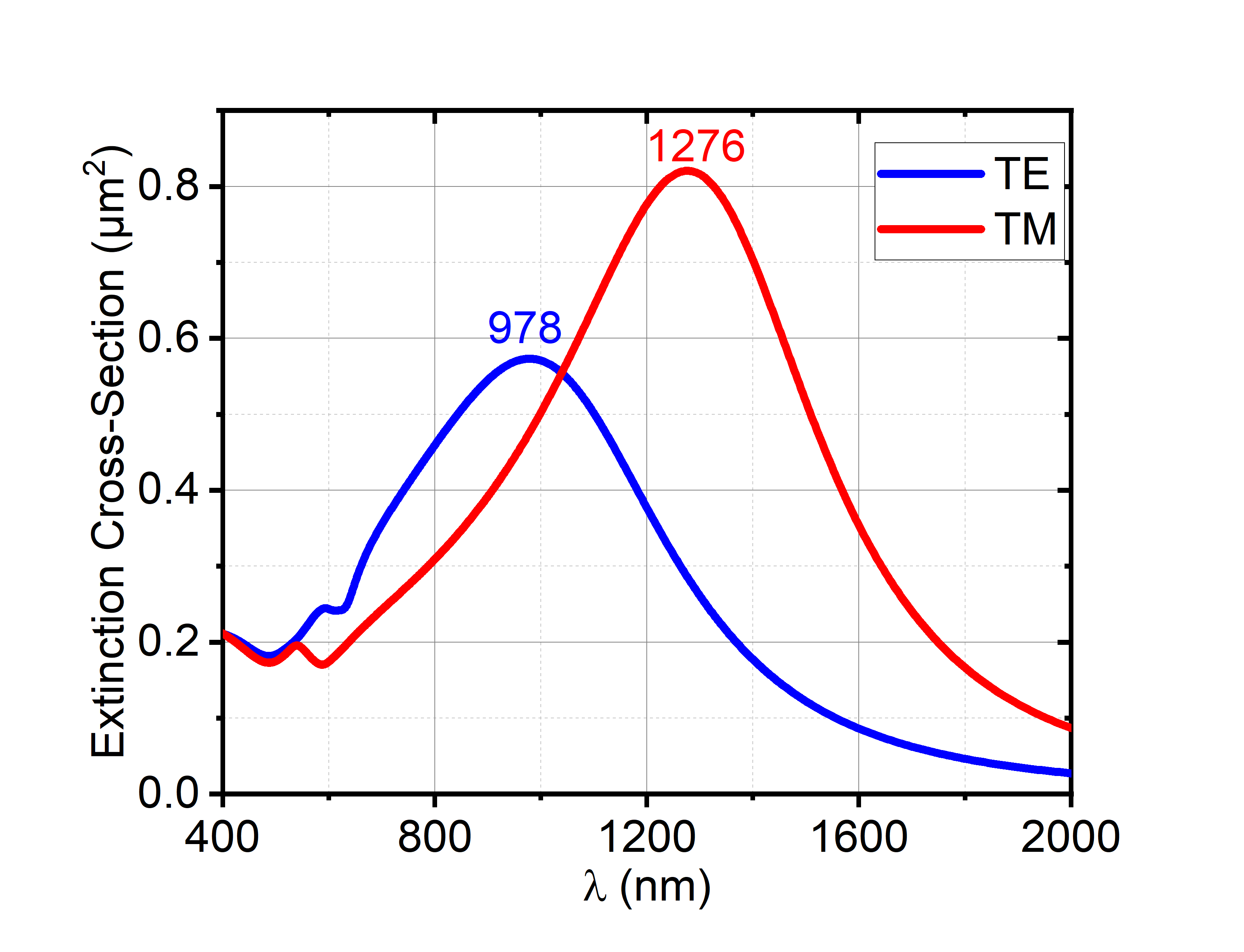}
  \caption{Simulated extinction cross-section versus the wavelength of a single isolated nanobar with geometric parameters $w_x$=\SI{400}{\nano \meter}, $w_y$=\SI{300}{\nano \meter}, $t$=\SI{50}{\nano \meter}. FDTD is used.}
  \label{fig:extinction_nanobar}
\end{figure}

\noindent
Figure~\ref{fig:extinction_nanobar} shows the simulated extinction cross-section versus the wavelength at normal incidence of a single isolated nanobar.
A peak is visible at the LSPR.

\newpage 
\subsection{Simulated linear transmission spectrum} 
\begin{figure}[h]
\centering
    \subfloat[TM, $P_x$=~540~nm]{\includegraphics[width=.5\textwidth]{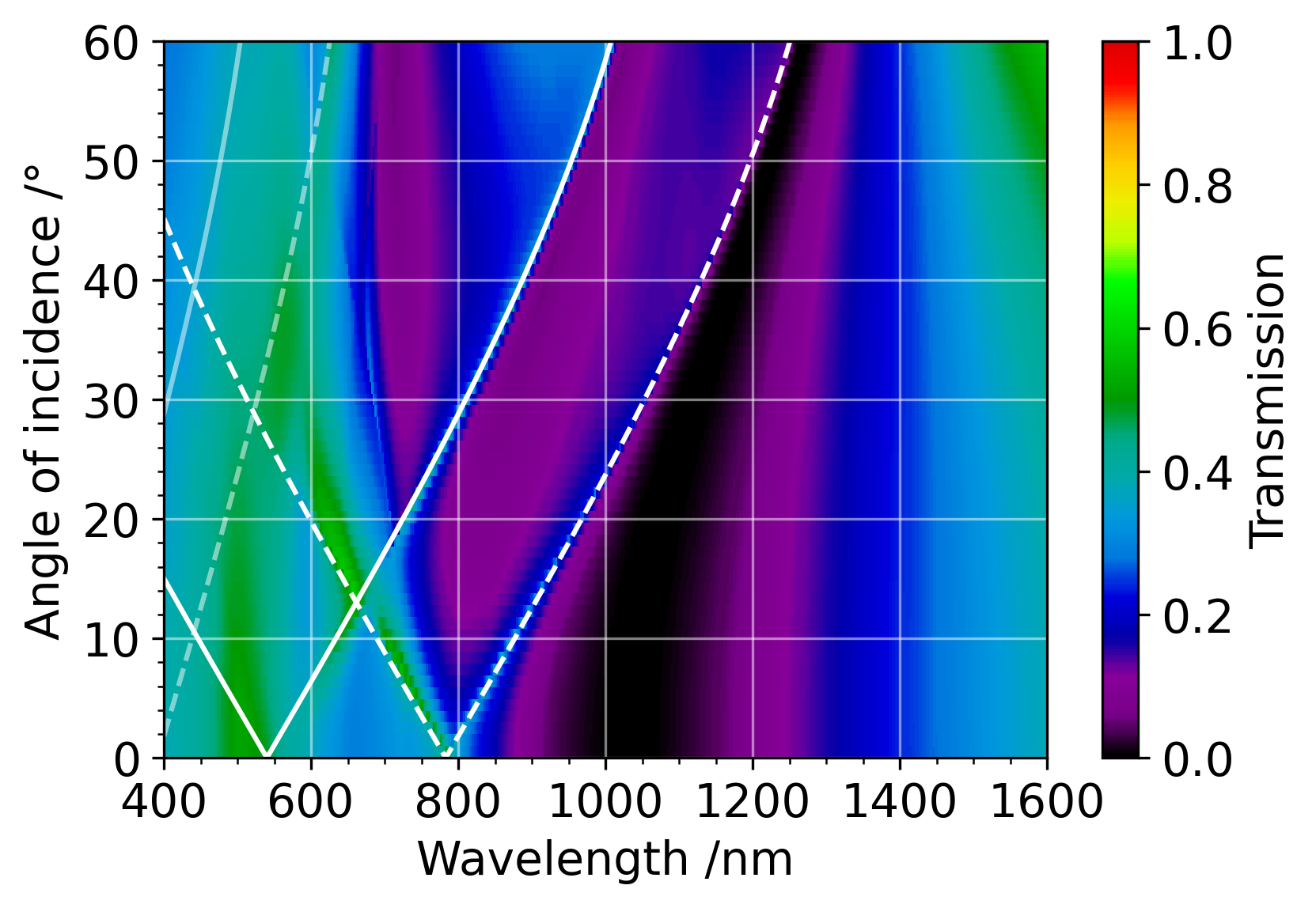}}
    \subfloat[TE, $P_x$=~540~nm]{\includegraphics[width=.5\textwidth]{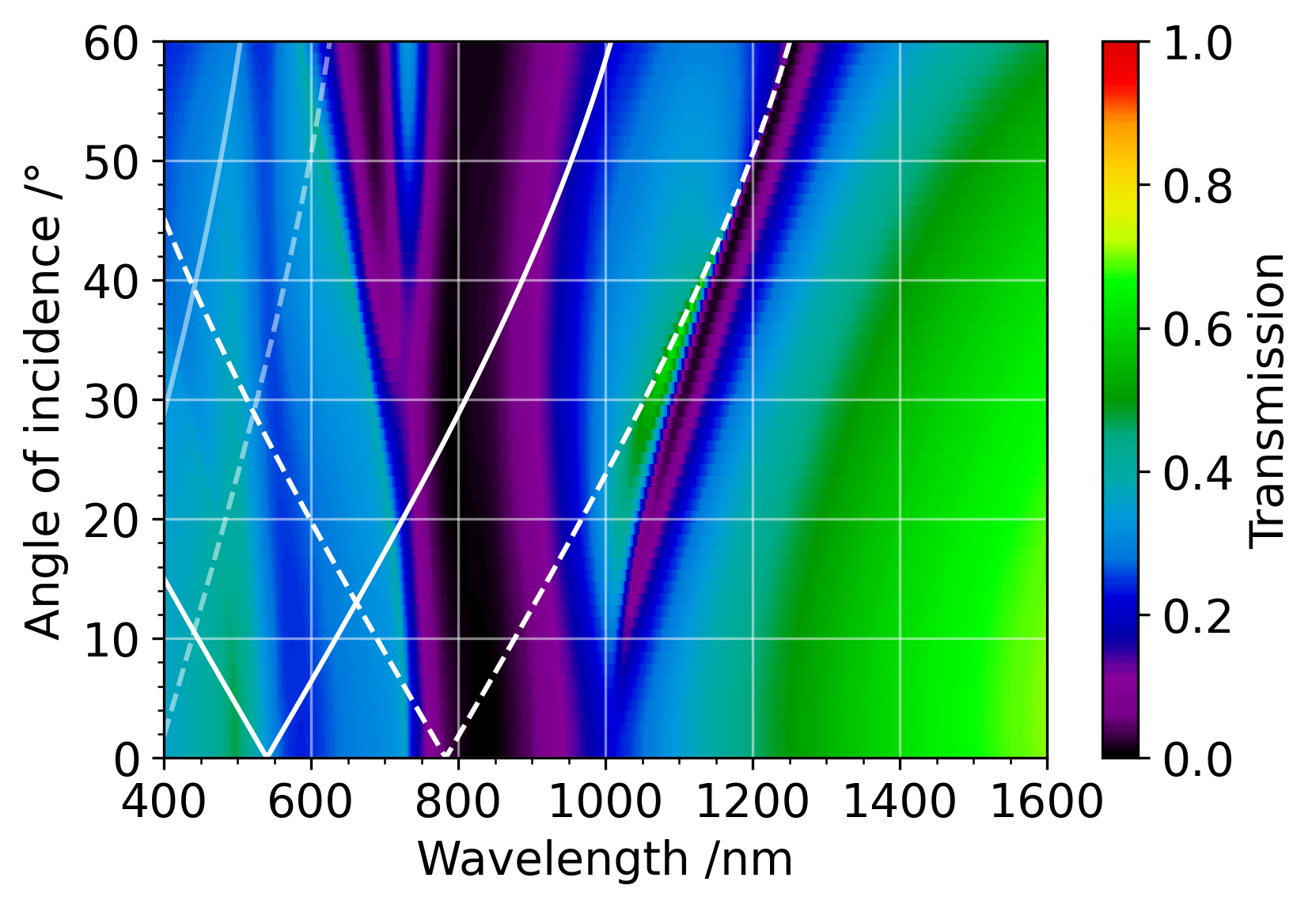}} \\
    \subfloat[TM, $P_x$=~1200~nm]{\includegraphics[width=.5\textwidth]{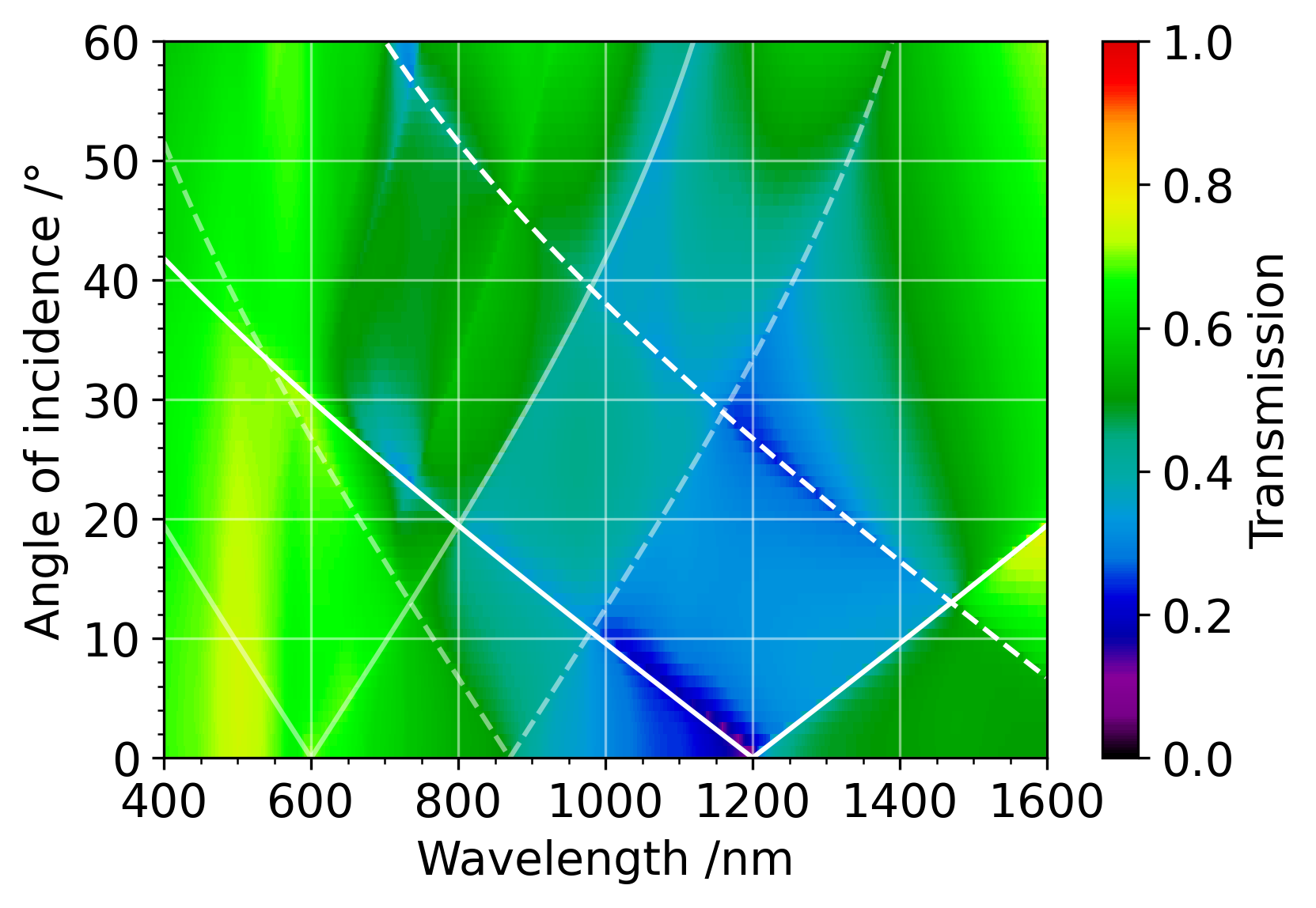}}
    \subfloat[TE, $P_x$=~1200~nm]{\includegraphics[width=.5\textwidth]{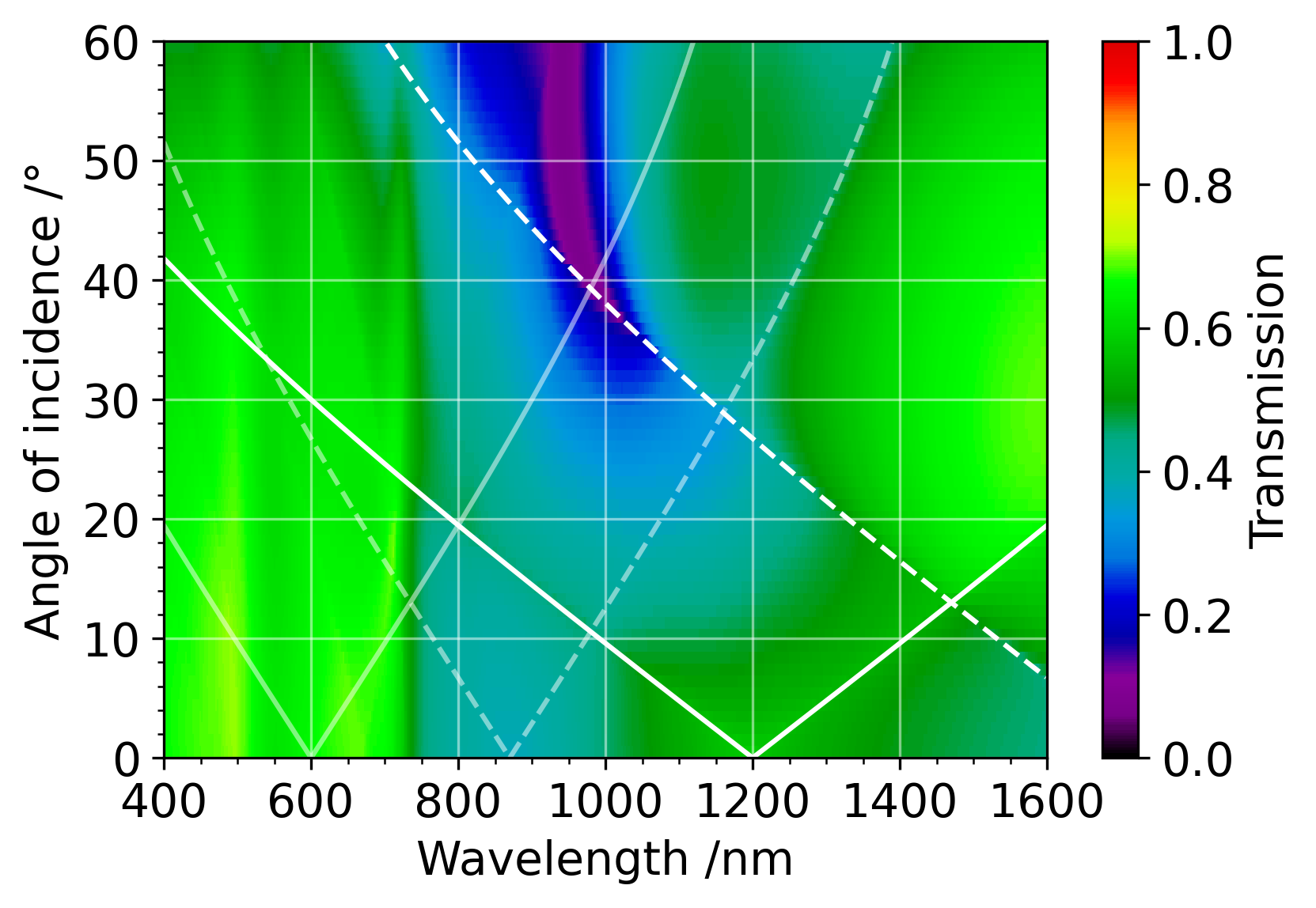}}
    \caption{Simulated linear transmission spectra for the 0\textsuperscript{th} order versus wavelength and incidence angle  for TM and TE linearly polarized input for $P_x$=\SI{540}{\nano \meter} and $P_x$=\SI{1200}{\nano \meter} nanobar structures, respectively. Inserted lines show the angle of the RAs, in (solid) air, and (dashed) fused silica. For the 1\textsuperscript{st} and (higher transparency) 2\textsuperscript{nd} diffraction order. RCWA is used.}
    \label{fig:linearspectra}
\end{figure}

\noindent
Figure~\ref{fig:linearspectra} shows the simulated linear transmission spectra for the shortest and longest fabricated periods as computed from the RCWA.
The broad transmission valley is independent from the periodicity, but changes with the polarization direction due to the nanobar shape.
The absolute transmission decreases for smaller periods due to the larger fill-factor.

\newpage \noindent
\subsection{Comparison between experiment and simulation} 
\begin{figure}[h]
\centering
    \subfloat[Experiment]{\includegraphics[width=.5\textwidth]{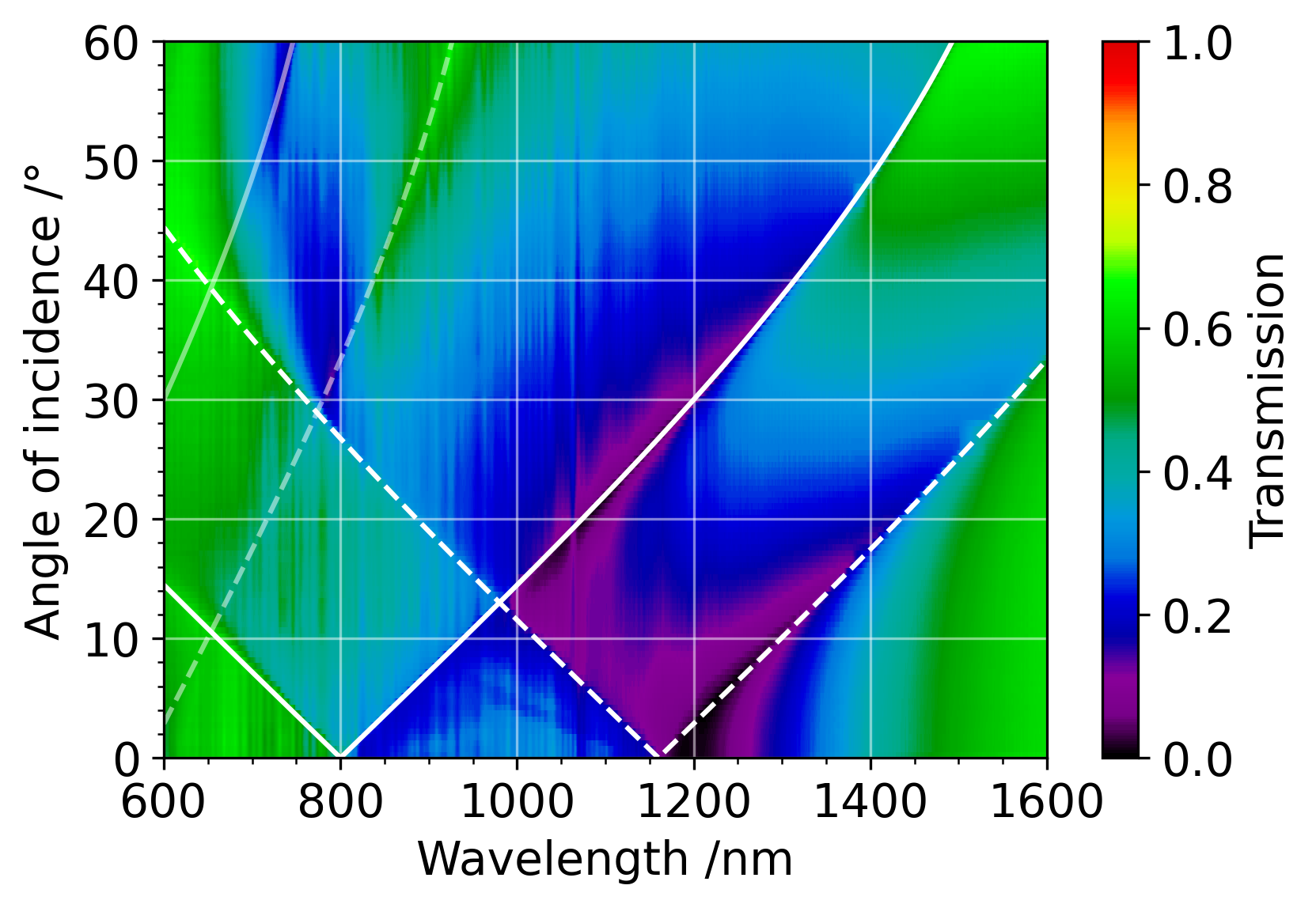}}
    \subfloat[RCWA simulation]{\includegraphics[width=.5\textwidth]{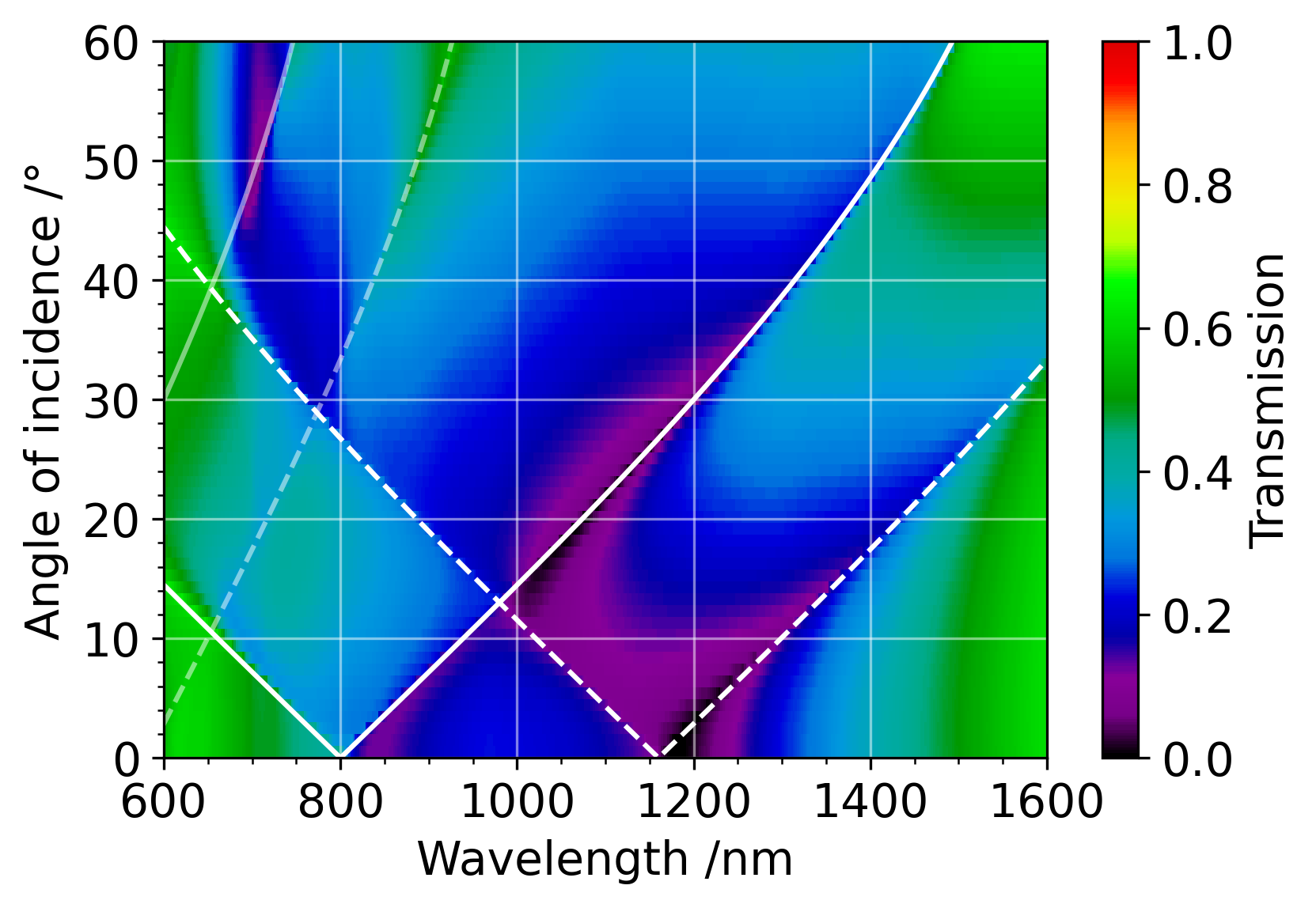}}
    \caption{Experimental and simulated (RCWA) linear transmission spectrum for TM polarized input for the $P_x$=\SI{800}{\nano \meter} nanobar structure. Inserted lines show the angle of the RAs for (solid) air, and (dashed) fused silica. For the 1\textsuperscript{st} and (higher transparency) 2\textsuperscript{nd} diffraction order.}
    \label{fig:linearspectra_compair}
\end{figure}

\noindent
Figure~\ref{fig:linearspectra_compair} shows the experimental (left panel) and theoretical (right panel) transmission spectra versus the incidence angle for a period of $P_x=\SI{800}{\nano\meter}$.
The simulated spectra stemming from Rigorous coupled-wave analysis (RCWA) match very well with the experimental results, regardless of $P_x$.

\newpage \noindent
\section{Second harmonic generation experiment} 
\begin{figure}[h]
\centering
\includegraphics[width=1\textwidth]{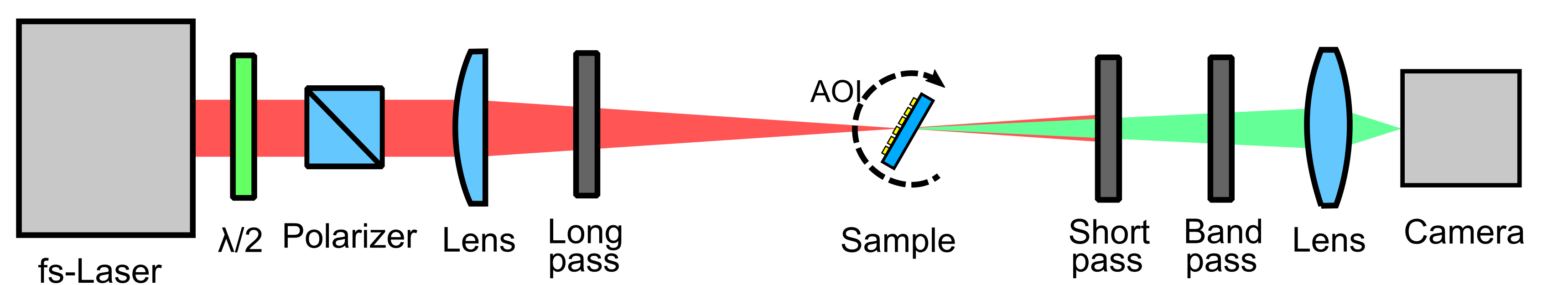}
  \caption{SHG characterization setup.}
  \label{fig:SHGsetup}
\end{figure}

\noindent
Figure~\ref{fig:SHGsetup} shows the SHG characterization setup.
An ultrafast laser (PHAROS-SP by Light Conversion) delivers \SI{200}{\femto\second} (FWHM) short pulses at a wavelength of \SI{1032}{\nano\meter} with a repetition rate of \SI{200}{\kilo\hertz} and an average power of \SI{150}{\milli\watt}.
The input power and linear polarization direction are adjusted via a half wave plate and a polarizer.
The beam is focused onto the structured area to a beam diameter of \SI{250}{\micro\meter} using a plano-convex \SI{400}{\milli\meter} focal length lens.
Before the sample, parasitic  optical radiation at shorter wavelengths is removed by a long pass filter (FELH0900 by Thorlabs).
During the measurements the sample is automatically rotated along the $y$-direction at the focus position.
We investigate the straight transmitted SHG light depending on the angle of incidence.
Behind the sample, the illumination light is seperated from the weak SH light with a short pass filter (FESH0800 by Thorlabs), which is highly transmissive (\SI{>95}{\percent}) from \SI{500}{\nano\meter} to \SI{789}{\nano\meter}.
In addition, a \SI{520}{\nano\meter} band pass filter (FBH520-40 by Thorlabs) with a FWHM of \SI{40}{\nano\meter} can be used to filter out ambient light.
To image the sample with its emitting signal, a \SI{100}{\milli\meter} bi-convex lens is placed at a distance of \SI{370}{\milli\meter} behind the sample.
An MgF$_2$ Rochon prism (RPM10 by Thorlabs) analyzer can be placed in front of the imaging lens to determine the polarization direction of the SHG.
The sample is imaged onto a CMOS camera (UI-1490SE-M-GL by iDS).
Once the background level is compensated for, the sum of the signal measured on each pixel overlapping with the observed SHG spot divided by the used exposure time is proportional to the average optical power.
The calibration factor of the camera to determine the impinging  SHG power was found in the following manner.
For the power calibration we inserted a BBO crystal in the setup instead of a sample.
The output power was then high enough to be measured with a calibrated photodiode (S121C by Thorlabs) at the position of the camera.
In the following step several calibrated neutral filters were inserted to reduce the power to the range of the second harmonic signal emitted from the metasurface.
Knowing the output power of the crystal and the optical density of the filters, it is then possible to connect the observed signal on the camera with the incident power at $2\omega$.
The ratio between the calculated input power and this camera value is the calibration factor.
From the captured images, the digital signal was integrated over the observed SHG spot, divided by the exposure time used and multiplied with the calibration value to get the optical power.
In the procedure, the linear behaviour of the camera was ensured for different exposure times and digital signal levels.
On the other side, to investigate the SHG spectrum, the output light is directly coupled into a compact spectrometer (Ocean HDX by Ocean Insight) with a back-thinned CCD. The image of spectrum shown in the main article are not subject to any processing, except for the subtraction of the constant background, the latter value being taken in region where no signal is measured.

The error bars of Fig.~7(a) take into account the measured small laser fluctuations and their effect on the SHG, the measured background noise and the resolution of the measurement devices. 
We find that the main error source is the thermal noise from the pump power detector. 
The \SI{3}{}--\SI{10}{\percent} error of the SHG power is scarcely  visible due to the logarithmic scale.
On the other hand, the error bars in Fig.~8 correspond to the difference between the SHG signal between positive and negative angles of incidence.

SHG from the pure fused silica surface is detectable at high angles of incidence for average powers above the damage threshold of the metasurface.
The maximum conversion efficiency is 3 order of magnitude lower than from the metasurface, therefore the susceptibility of the glass surface is neglected.

\newpage
\section{Second harmonic generation simulation}
\subsection*{Nonlinear inverse scattering technique}

Let us consider two electrical currents $\mathbf{J}_1$ and $\mathbf{J}_2$, sources respectively of two electric fields $\mathbf{E}_1$ and $\mathbf{E}_2$. 
When both dielectric and magnetic permittivity tensors are symmetric, the Lorentz reciprocity theorem in the harmonic regime (fields proportional to $e^{i\omega^\prime t}$) states
\begin{equation} \label{eq:Lorentz}
    \int{\mathbf{J}_2(\mathbf{r}^{\prime\prime},\omega^\prime)\cdot \mathbf{E}_1(\mathbf{r}^{\prime\prime},\omega^\prime)dV^{\prime\prime}}= \int{\mathbf{J}_1(\mathbf{r}^\prime,\omega^\prime)\cdot \mathbf{E}_2(\mathbf{r}^\prime,\omega^\prime)dV^\prime}.
\end{equation}
We want to apply Equation~\ref{eq:Lorentz} to predict the SH emission from a periodic metallic metasurface. Following \cite{Obrien:2015}, we consider $\mathbf{J}_1$ to be originated from the nonlinear polarization at $\omega^\prime=2\omega$. We thus set $\mathbf{J}_1=2i\omega \mathbf{P}^\mathrm{NL}$. 
The other source $\mathbf{J}_2$ is instead a localized dipole $\mathbf{J}_2(\mathbf{r}^{\prime\prime}) = J_0 \delta(\mathbf{r}^{\prime\prime}-\mathbf{r})\hat{j}$, used to sample in each spatial point $\mathbf{r}$ the $j$-th component of the field $\mathbf{E}_1$ emitted from $\mathbf{J}_1$. Equation~\ref{eq:Lorentz} then provides \cite{Roke:2004}
\begin{equation} \label{eq:SHG_emitted_dipole}
     \mathbf{E}_\mathrm{emission}(\mathbf{r},2\omega) \cdot \hat{j}= \frac{ 2i\omega 
     }{J_0} \int{\mathbf{P}^\mathrm{NL}(\mathbf{r^\prime},2\omega)\cdot \mathbf{E}_\mathrm{dipole}(\mathbf{r}^\prime,2\omega)dV^\prime},
\end{equation}
where we set $\mathbf{E}_1=\mathbf{E}_\mathrm{emission}$ and  $\mathbf{E}_2=\mathbf{E}_\mathrm{dipole}$. The field $\mathbf{E}_\mathrm{emission}$ is the second harmonic emitted from the metasurface, whereas $\mathbf{E}_\mathrm{dipole}$ is the field on the metasurface induced by the elemental dipole $\mathbf{J}_2$ placed in $\mathbf{r}$. 
As well known, the presence of nonlinearity breaks the reciprocity of the system \cite{Caloz:2018}, seemingly invalidating Equations~\ref{eq:Lorentz} and \ref{eq:SHG_emitted_dipole}. The point is that the derivation presented above assumes a fixed nonlinear polarization, thus treating it as a known forcing term. 
This approach holds valid in the non-depleted regime, where the number of photons in the fundamental beam is not significantly reduced by the SHG.
Nonetheless, the nonlinear inverse scattering technique has been also generalized to more general conditions, including losses and saturation \cite{GomezDiaz}. \\
The last ingredient is then finding the correct form to model the nonlinear polarization. In our case the main contribution comes from the surface nonlinearity at the gold interfaces. Accounting for the isotropic response of gold, the associated elements of the nonlinear tensors are $\chi^{(2)}_{nnn}=4.42\times 10^{-22}$m$^2$/V, $\chi^{(2)}_{ntt}=4.94\times 10^{-21}$m$^2$/V and $\chi^{(2)}_{ttn}=1.77\times 10^{-21}$m$^2$/V \cite{Wang:2009}, where the subscripts $t$ and $n$ refer to the components of the electric field tangential and normal to the interface, respectively. \\
Equation~\ref{eq:SHG_emitted_dipole} is already apt to describe the second harmonic emitted from our metasurface. But given the periodicity of our structure, it is more convenient from a numerical point of view to compute the electric field inside the integral by considering a plane wave excitation \cite{Obrien:2015}. 
The electromagnetic field radiated from a dipole tends to a plane wave when the observation point is far enough from the source. 
Thus, we suppose the point $\mathbf{r}$ to be far enough from the metasurface, corresponding to compute $\mathbf{E}_\mathrm{emission}$ in the far field. 
The Green function $G$ in free space for the magnetic potential vector $\mathbf{A}$ is the spherical wave $G(\mathbf{r}-\mathbf{r}^{\prime})=\frac{\mu}{4\pi|\mathbf{r}-\mathbf{r}^{\prime}|}e^{-ik|\mathbf{r}-\mathbf{r}^{\prime}|}$, where $k$ is the wavevector in air. 
In the Lorenz gauge the electric field can be derived from the magnetic potential vector $\mathbf{A}$ using $\mathbf{E}=-2i\omega \mathbf{A} - i/\left( 2\omega\mu\epsilon\right)\nabla\nabla\cdot \mathbf{A}$, in turn providing in the far field
\begin{equation} \label{eq:dipole_farfield}
 \mathbf{E}_\mathrm{dipole} = \frac{2i\omega\mu J_0}{4\pi r} e^{-ikr} \sin\theta,
\end{equation}
where $r$ is the distance from the source and $\theta$ is the angle between the dipole axis and the emission direction. 
Next, we need to find the relationship between the amplitudes of the dipole $J_0$ and the impinging plane wave $a_0$. 
Such connection depends on the relative orientation between the dipole and the interface, i.e., the emitted polarization $\hat{j}$.
For TE outputs ($\hat{j}$ parallel to $\hat{y}$) the dipole stays parallel to the interface regardless of the incidence angle $\theta_{AOI}$. 
Thus in Equation~\ref{eq:dipole_farfield} it is always $\theta=\pi/2$, in turn providing
\begin{equation} \label{eq:dipole_PW_TE}
    a_0^{(TE)} = \frac{2i\omega\mu J_0}{4\pi r}.
\end{equation}
For TM outputs it is $\hat{j}= \hat{x} =\cos\theta_{AOI} \hat{t} + \sin\theta_{AOI} \hat{n}$, where $\hat{t}$ and
$\hat{n}$ are the unit vectors parallel and normal to the interface. The component of the dipole along $\hat{t}$ behaves like the TE component and follows Equation~\ref{eq:dipole_PW_TE} as well. 
The normal component of the dipole is equivalent to $\theta=\pi/2$, thus radiative coupling with the metasurface is forbidden.
Remembering that $\theta=\pi/2-\theta_{AOI}$ and that the effective dipole amplitude is $J_0 \cos\theta_{AOI}$ we find
\begin{equation} \label{eq:dipole_PW_TM}
    a_0^{(TM)} = \frac{2i\omega\mu J_0}{4\pi r} \cos\theta_{AOI}.
\end{equation}
Equation~\ref{eq:dipole_PW_TM} can also be found more directly from Equation~\ref{eq:dipole_farfield} by considering the metasurface being rotated around the constant linear dipole of amplitude $J_0$ and accounting for the relation between $\theta$ and $\theta_{AOI}$.
All the above polarization-dependent results can be gathered together defining the diagonal matrix in the laboratory framework $xyz$
\begin{equation}
    \mathbf{\Upsilon} = \frac{1}{a_0} \left( \begin{array}{ccc}
        \cos(\theta_{AOI}) & 0 & 0\\
        0 & 1 & 0 \\
        0 & 0 & 0
    \end{array} \right) .
\end{equation}
Substituting back to Equation~\ref{eq:SHG_emitted_dipole} we find
\begin{equation} \label{eq:SHG_emitted_pw}
     \mathbf{E}_\mathrm{emission}(\mathbf{r},2\omega) \cdot \hat{j}= -\frac{ 4\mu\omega^2  
     }{4\pi r} \int{\mathbf{P}^\mathrm{NL}(\mathbf{r^\prime},2\omega)\cdot \left[\mathbf{\Upsilon} \cdot \mathbf{E}_\mathrm{planewave}(\mathbf{r}^\prime,2\omega) \right] dV^\prime}.
\end{equation}
The new field $\mathbf{E}_\mathrm{planewave}$ is the field on the surface of the metasurface when it is illuminated with a plane wave of amplitude $a_0$, frequency $2\omega$ and wavevector $\mathbf{k}=\cos\theta_{AOI}\ \hat{z}  + \sin\theta_{AOI}\ \hat{x} $. With respect to the original formula derived in \cite{Obrien:2015}, we have an additional anisotropic factor modelled by the matrix $\mathbf{\Upsilon}$ accounting for the skewed incidence angle on the sample.

\subsection*{Green function approach}

The nonlinear inverse scattering technique described in the previous paragraph can be interpreted directly in terms of antenna theory. 
The crucial point is that we are assuming the nonlinear polarization $\mathbf{P}^\mathrm{NL}$ to be fixed and determined by the the interaction of the fundamental beam with the metasurface in the linear regime. 
For a distribution $\mathbf{J}$ of currents, the emitted electric field is the superposition integral between $\mathbf{J}$ and the dyadic Green function $\mathbf{\Gamma}= -i\omega \left[ \mathbf{I} + k^{-2}\nabla\nabla\right] G_\mathrm{int} $ \cite{Arnoldus:2001}, where $G_\mathrm{int}$ is the Green function accounting for the presence of the interface glass-air and the metallic nanostructures. For nonlinear thin materials we find
\begin{equation}
    \label{eq:SHG_antenna_theory}
     \mathbf{E}_\mathrm{emission}(\mathbf{r};2\omega) = -2i\mu\omega 
      \int{\mathbf{\Gamma}(\mathbf{r}^\prime,\mathbf{r};2\omega) \cdot \mathbf{P}^\mathrm{NL}(\mathbf{r^\prime};2\omega) dV^\prime}.
\end{equation}
The equivalence between Equation~\ref{eq:SHG_emitted_dipole} and Equation~\ref{eq:SHG_antenna_theory} can be established by accounting for the symmetry properties of the tensorial Green function stemming from the reciprocity theorem. 
As a matter of fact, reciprocity imposes on the dyadic Green function the constraint $\Gamma_{mn}(\mathbf{r^\prime},\mathbf{r})=\Gamma_{nm}(\mathbf{r},\mathbf{r^\prime})$: exchanging the position of the field probe and of the emitting dipole does not induce variations in the electromagnetic response, if  the respective polarization directions are simultaneously interchanged.
When the $j$-th component of $\mathbf{E}_\mathrm{emission}$ is calculated, the elements of the dyadic product $\sum_{k}\Gamma_{jk}P^\mathrm{NL}_k$ are selected. 
Accordingly, in Equation~\ref{eq:SHG_emitted_dipole} the emitting current is taken parallel to the direction $\hat{j}$, thus confirming the swapping of the indices in the Green function.
From a physical point of view, this alternative interpretation based upon the Green function  stresses out that Equation~\ref{eq:SHG_emitted_pw} accounts for the exact Green function $G_\mathrm{int}$ of the structure, including both the reflection from the glass-air interface and the interaction with the metallic nano-structures. 
Direct usage of Equation~\ref{eq:SHG_antenna_theory} requires to find numerically the far field expression for $\mathbf{\Gamma}$ in each point $\mathbf{r}^\prime$ on the elemental unit. 
In the nonlinear inverse scattering approach discussed in the previous section, the numerical effort instead consists in the solution of the associated linear problem for only the direction of interest. 
Thus, in our case the nonlinear inverse scattering technique minimizes the demand of computational time.


\subsection*{Evaluation of the effective metasurface nonlinearity}

The unidirectional second harmonic generation in the Type I case is macroscopically described by \cite{Boyd:2008}
\begin{gather}
 \frac{\partial E_{2\omega}}{\partial z} = - \frac{i\omega}{cn(2\omega)} d_\mathrm{eff} E^2(\omega) e^{i\Delta k z}  \label{eq:SHG_FF},\\   
 \frac{\partial E_{\omega}}{\partial z} = - \frac{i\omega}{cn(\omega)} d^*_\mathrm{eff} E(2\omega) E^*(\omega) e^{-i\Delta k z}, \label{eq:SHG_SH}
\end{gather}
where $d_\mathrm{eff}$ is the effective second order nonlinearity for the given input and output polarizations, $*$ stands for complex conjugation, and $\Delta k$ is the phase mismatch. In the case of a metasurface placed in $z=0$, we can assume a delta-Dirac functional form for the nonlinearity seen on a macroscopic scale, $d_\mathrm{eff}=d_\mathrm{surf} \delta(z)$. Accounting for the metasurface thickness $h=50~$nm, we can set $d_\mathrm{surf}= h d_\mathrm{meta}$ where $d_\mathrm{meta}$ is defined to account for the overall average longitudinal response of the metasurface. Under the no-depletion hypothesis, integration of Equation~\ref{eq:SHG_FF} along $z$  yields for $z>0$
\begin{equation}
    E(2\omega) =- \frac{i k_0 h}{ n(2\omega)} d_\mathrm{meta} E^2(z=0,\omega).
\end{equation}
By rewriting the electric fields in term of the respective intensity $I=0.5cn\epsilon_0|E|^2$
\begin{equation} \label{eq:effective_nonlinearity}
    |d_\mathrm{meta}| =  \frac{\kappa(\omega)}{h} \left( \frac{I(2\omega)} {I_\mathrm{surf}^2(\omega)} \right)^{1/2},
\end{equation}
where we defined $\kappa(\omega)=\sqrt{ \frac{c\epsilon_0 n(2\omega) n^2(\omega)}{2k_0^2}}$, $I(2\omega)$ is the second harmonic intensity emitted in the forward direction and $I_\mathrm{surf}(\omega)$ is the intensity effectively coupled to the metasurface. 
Given that the refractive index of a metasurface is not well defined, for simplicity we use a refractive index of $n(\omega)=n(2\omega)$=\SI{1}{} for both frequencies.\\
From an experimental point of view, the next step is to connect the intensity $I_\mathrm{surf}$ to the impinging intensity $I_0$. The connection is not straightforward, as shown in the seminal paper by Chen concerning the surface-enhancement of SHG \cite{Chen:1983}. 
Given we are impinging from the air side, in a first approximation the effective field on the metallic nano-structure is $E_\mathrm{inc}[1+R(\theta_{AOI};j)]$, where $R(\theta_{AOI};j)$ is the Fresnel coefficient for plane waves polarized along $\hat{j}$ at the interface air-glass. 
In this approximation we find $I_\mathrm{surf}=|1+R(\theta_{AOI};j)|^2I_\mathrm{inc}$. Equation~\ref{eq:effective_nonlinearity} turns into 
\begin{equation} \label{eq:effective_nonlinearity_exp_power}
    |d_\mathrm{meta}| =  \frac{\kappa(\omega)}{h |1+R(\theta_{AOI};j)|^2} \left( \frac{I(2\omega)} {I^2_\mathrm{inc}(\omega)} \right)^{1/2}.
\end{equation}The previous approach is valid in the limit of plane waves and monochromatic waves. For a proper comparison with our experiments, we need to account for the pulse duration, beam width and repetition rate. The instantaneous intensity can be factorized out as $i_\mathrm{inc}(t)=i_\mathrm{peak} u(t) f^2(x,y)$. We then set $f(x,y)=e^{-(x^2+y^2)/w_\mathrm{inc}^2}$,  $u(t)=e^{-2t^2/\tau^2}$ and $i_\mathrm{peak}=2\sqrt{2}\mathcal{E}/\left(\pi^{3/2} \tau w^2_\mathrm{inc} \right)$, where $w_\mathrm{inc}$ is the beam width at the metasurface, $\tau$ is the ($1/e^2$) pulse duration, and $\mathcal{E}$ is the pulse energy.  Strictly following the concept of root mean square, the equivalent monochromatic plane wave intensity $E_\mathrm{eq}(\omega)$ is found by a spatio-temporal average across the electromagnetic pulse 
\begin{equation}
    \frac{P_\mathrm{eq}}{d_\mathrm{eff}} = E_\mathrm{eq}^2(\omega) \equiv \frac{2 i_\mathrm{peak}}{c n \epsilon_0} \frac{1}{\tau}\frac{1}{w_\mathrm{inc}^2}\int_{pulse}{u(t) dt}\int_{beam}{f^2(\mathbf{r})dS}.
\end{equation}
Limiting the integral to the $1/e^2$ width, the equivalent square field is
\begin{equation} \label{eq:effective_square_field}
    E^2_\mathrm{eq}(\omega) = \frac{2}{c n \epsilon_0} \frac{\mathcal{E}}{\tau w^2_\mathrm{inc} }  \frac{\text{erf}(\tau)\  \text{erf}^2 (w_\mathrm{inc})}{\tau w^2_\mathrm{inc} } \approx \frac{2}{c n \epsilon_0} \left(\frac{2}{\pi}\right)^{3/2} \frac{\mathcal{E}}{\tau w^2_\mathrm{inc} }, 
\end{equation}
where $\text{erf}(x) = \left(2/\sqrt{\pi}\right) \int_0^x \exp{\left(-t^2\right)dt}$ is the standard error function, and $\text{erf}(x)\approx \left(2/\sqrt{\pi}\right)x$ for small enough $x$.
For a train of pulses, the average power $\mathcal{P}$ is related to the pulse energy $\mathcal{E}$ via the repetition rate $f_\mathrm{rep}$, $\mathcal{P}=\mathcal{E}f_\mathrm{rep}$. 
From the experimental side, the known quantities are the emitted average second harmonic power $\mathcal{P}_{measured}(2\omega)$ and the measured beam width $w_{measured}$, the latter being measured with the camera at large enough generation. 
Substituting Equation~\ref{eq:effective_square_field} into Equation~\ref{eq:effective_nonlinearity_exp_power}, we finally find the effective nonlinear coefficient of the metasurface
\begin{equation} \label{eq:effective_nonlinearity_exp_power_final}
    |d_\mathrm{meta}| =  \frac{\kappa(\omega)}{h |1+R(\theta_{AOI};j)|^2} \left(\frac{\pi}{2} \right)^{3/2} \frac{ w^2_\mathrm{inc} \sqrt{ \tau f_\mathrm{rep}}}{{w_{measured}}}  \left( \frac{\mathcal{P}_{measured}(2\omega)} {\mathcal{P}^2_\mathrm{inc}(\omega)} \right)^{1/2}.
\end{equation}

Using a pump pulse with a wavelength of \SI{1032}{\nano\meter}, \mbox{$\tau=\SI{200}{\femto\second}$},
\mbox{$f_{rep}=\SI{200}{\kilo\hertz}$}, 
\mbox{$P_{inc}=\SI{180}{\milli\watt}$},
\mbox{$\omega_{inc}=\SI{125}{\micro\meter}$} and
\mbox{$R(\SI{20.5}{\degree},TM)=\SI{0.166}{}$} leads to a SHG signal with \mbox{~$\omega_{measured}=\SI{40}{\micro\meter}$} and \linebreak \mbox{$P_{measured}=\SI{190}{\pico\watt}$}. 
Therefore, the $h=\SI{50}{\nano\meter}$ thick metasurface has a second order nonlinearity $d_\mathrm{meta}$ of \SI{1.0}{\pico\meter\per\volt}.

\end{document}